\documentclass[12pt]{article}

\usepackage{amsmath,amssymb,bm}

\newtheorem{theorem}{Theorem}[section]
\newtheorem{corollary}[theorem]{Corollary}
\newtheorem{lemma}[theorem]{Lemma}

\newcommand{\B}{\ensuremath{{\cal B}}}
\newcommand{\K}{\ensuremath{{\cal K}}}
\newcommand{\qed}{~ \rule{0.8ex}{1.6ex}\smallskip}

\title{Gapless Excitation above a Domain Wall Ground State in a Flat-Band Hubbard Model}

\author{Makoto Homma and Chigak Itoi \vspace*{0.5em}\\ {\it Department of Physics, Science and Technology,}\\ {\it Nihon University, Tokyo, Japan}}

\begin{document}

\maketitle

\begin{abstract}
 We construct a set of exact ground states with a localized ferromagnetic domain wall and with an extended spiral structure in a deformed flat-band Hubbard model in arbitrary dimensions. We show the uniqueness of the ground state for the half-filled lowest band in a fixed magnetization subspace. The ground states with these structures are degenerate with all-spin-up or all-spin-down states under the open boundary condition. We represent a spin one-point function in terms of local electron number density, and find the domain wall structure in our model. We show the existence of gapless excitations above a domain wall ground state in dimensions higher than one. On the other hand, under the periodic boundary condition, the ground state is the all-spin-up or all-spin-down state. We show that the spin-wave excitation above the all-spin-up or -down state has an energy gap because of the anisotropy.
\smallskip

\noindent%
{\bf keywords}:
 ferromagnetic domain wall, spiral state, flat-band Hubbard model, exact
 solution, quantum effect, spin-wave, gapless excitation

\end{abstract}

%\tableofcontents

\section{Introduction}

Domain structures are observed universally
in many ferromagnetic systems. If a system has a translational symmetry,
this symmetry is broken spontaneously by domains.
In classical spin systems, universal
properties of the domain wall have been studied extensively. For example in
the Ising model on the cubic lattice, Dobrushin proved that a horizontal
domain wall is stable against the thermal fluctuations at sufficiently
low temperatures \cite{D}. This structure in the Ising model is also
preserved under quantum perturbations. Borgs, Chayes and Fr\"ohlich
proved that the horizontal domain wall on the
$d$-dimensional hyper cubic lattice is stable also against
weak quantum perturbations at sufficiently low temperatures
for $d \geq 3$ \cite{BCF}. On the other hand, a diagonal domain wall
structure is unstable in the Ising model,
since some local operators can deform the diagonal domain wall
state to many other ground states without loss of energy.
In the ferromagnetic
XXZ model in dimensions higher than one, however, no local operator can
deform the diagonal domain wall ground state to other ground states by
the exchange interaction. This fact suggests that the diagonal domain
structures are stable at sufficiently low temperatures in sufficiently
high dimensions, even though there is no proof.
Alcaraz, Salinas and Wreszinski construct a set of ground state with
diagonal domain wall structure in the XXZ model with a critical boundary
field in arbitrary dimensions for an arbitrary spin \cite{ASW}.
Gottstein and Werner clarified the structure of ground states
in the one-dimensional XXZ model with an infinite-volume \cite{GW}.
Koma and Nachtergaele proved that there is an energy gap above any
ground states in the one-dimensional XXZ model \cite{KN}.
They also showed an interesting result that there exists a gapless excitation above the
domain wall ground state in the XXZ model in two dimensions \cite{KN2d}.
Matsui extended this theorem to the XXZ model in arbitrary dimensions
higher than one \cite{M}. Bolina, Contucci, Nachtergaele and Starr
gave more precise bound for the gapless excitation above the diagonal domain wall
ground state in the XXZ model \cite{BCNS}.
Bach and Macris evaluate a spin one-point function in the
domain wall state of the one-dimensional XXZ model
by a rigorous perturbation method \cite{BM}.
Datta and Kennedy also discussed the
existence of a domain wall in one-dimensional XXZ models by another
rigorous perturbation method \cite{DK}.
They show that the exchange interaction destroys
the domain wall in the antiferromagnetic model,
while the domain wall exists in the ferromagnetic model at zero temperature.
The role of the quantum effects
should be studied more in many other models.

Recently, a deformed flat-band Hubbard model with an exact domain wall 
ground state was proposed \cite{HI,HI2}. The purpose of this paper is to study excitations
above the domain wall ground state
in this model and clarify whether or not, this model has the same spectra as
in the XXZ model.
The flat-band Hubbard model was proposed as a lattice electron model
with a ferromagnetic ground state.
Some remarkable results for ferromagnetic ground states have been
obtained in this class of models. Mielke and Tasaki have independently
shown that the ground state gives saturated ferromagnetism in
a class of many-electron models on a lattice,
which are called flat-band Hubbard models \cite{Mielke, T0}.
Nishino, Goda and Kusakabe extended their result to more general models \cite{K}.
Tasaki proved also the stability of the saturated ferromagnetism against a
perturbation which bends the electron band \cite{T}. Tanaka and Ueda
have shown the stability of the saturated ferromagnetism in a more
complicated two-dimensional model in Mielke's class \cite{TU}.
Tasaki has studied the energy of the spin-wave excitations in the
flat-band Hubbard model \cite{Tasaki}. He has shown that
the dispersion of the one-magnon
excitation is non-singular in the flat-band Hubbard model,
contrary to the Nagaoka ferromagnetism.
 The flat-band ferromagnetism is believed to be stable
against a small perturbation or the change of the electron number
density\cite{MT}. 
Unlike the ferromagnetic quantum spin model, we expect strong
quantum effects in the ferromagnetic ground state of the electrons on
the lattice.
The fermion statistics and fully polarized spin configuration imply that this state
is microscopically entangled with respect to the electron site
configuration. Therefore, the calculations of the ground
state expectation value
become more complicated than those in the XXZ model in which the ground state
can be written in a product state.

Here, we deform a flat-band Hubbard model by a complex anisotropy
parameter $q$. The SU(2) spin rotation symmetry in the original
flat-band model is reduced to U(1) symmetry in our deformed
model. First, we study our model under an open boundary condition. The
anisotropy $|q| \neq 1$ leads to a localized domain wall with finite
width. The domain structure is characterized in terms of the local order
parameter $\langle S_{x} ^{(3)} \rangle$, which represents the third
component of the localized spin at site $x$. This local order parameter
takes the same sign within one domain. The domain wall center is a set
of sites $x$ defined by zeros of the local order parameter
$\langle S_{x} ^{(3)} \rangle = 0$. We show the uniqueness of the ground
state with a fixed magnetization in a half-filled electron number in the
lowest energy band. We represent $\langle S_{x} ^{(3)} \rangle$ in terms
of the local electron density $\langle n_{x} \rangle$, and show the
profile of the ferromagnetic domain wall. We study the low energy
excitations in this model. We show that there exists a gapless
excitation above the domain wall ground state. This excited state is
constructed by acting a local operator near the domain wall on the
ground state. We discuss reliability of our results in the
infinite-volume limit, although we present our result with a finite
system size. This property of the domain wall ground state is similar to
the gapless excitation above the domain wall ground state in the XXZ
model as well. Next, we study the model under the periodic boundary
condition. In this case, either all-spin-up or all-spin-down state is
allowed as a ground state. We show that a spin-wave excitation above the
all-spin-up ground state has an energy gap because of the
anisotropy. This property is similar to the Ising gap in the
ferromagnetic XXZ model.

This paper is organized as follows. In section \ref{sec:def}, we define
a deformed flat-band Hubbard model on a decorated $d$-dimensional
integer lattice. In section \ref{sec:GS}, we construct a set of ground
states and prove the uniqueness of the ground state in a subspace
with each fixed magnetization. The domain wall structure is shown in
terms of the spin one-point function. We also obtain a representation
for the spin correlation function.
In section \ref{sec:Gapless}, we show the existence of gapless
excitation above the domain wall ground state in a sufficiently large
system size. An upper bound on the excitation energy
is given in Theorem \ref{th:gapless2} and Corollary \ref{th:gapless2}.
In section \ref{sec:SW}, we consider our model under the periodic boundary
condition. We estimate an energy gap of the spin-wave excitation above
the all-spin-up ground state. 
Finally, we summarize our results in section \ref{sec:summary}.

\section{Definition of the Model\label{sec:def}}

The Hubbard model is a model which represents a many-electron system on
an arbitrary lattice. In this section, we define a $d$-dimensional
deformed flat-band Hubbard model illustrating its physical
meaning. Our model is a generalization of the Tasaki model given in\cite{T0}.

\subsection{Lattice}

\begin{figure}[htbp]
 \begin{center}
  \setlength{\unitlength}{1mm}
  \begin{picture}(60,60)(-30,-30)
   \put(-20,-30){\circle*{2}}
   \put(-20,-10){\circle*{2}}
   \put(-20,10){\circle*{2}}
   \put(-20,30){\circle*{2}}
   \put(0,-30){\circle*{2}}
   \put(0,-10){\circle*{2}}
   \put(0,10){\circle*{2}}
   \put(0,30){\circle*{2}}
   \put(20,-30){\circle*{2}}
   \put(20,-10){\circle*{2}}
   \put(20,10){\circle*{2}}
   \put(20,30){\circle*{2}}
   \put(-30,-20){\circle*{2}}
   \put(-10,-20){\circle*{2}}
   \put(10,-20){\circle*{2}}
   \put(30,-20){\circle*{2}}
   \put(-30,0){\circle*{2}}
   \put(-10,0){\circle*{2}}
   \put(10,0){\circle*{2}}
   \put(30,0){\circle*{2}}
   \put(-30,20){\circle*{2}}
   \put(-10,20){\circle*{2}}
   \put(10,20){\circle*{2}}
   \put(30,20){\circle*{2}}
   \put(-20,-20){\circle{3}}
   \put(-20,0){\circle{3}}
   \put(-20,20){\circle{3}}
   \put(0,-20){\circle{3}}
   \put(0,0){\circle{3}}
   \put(0,20){\circle{3}}
   \put(20,-20){\circle{3}}
   \put(20,0){\circle{3}}
   \put(20,20){\circle{3}}
   \put(-20,-30){\line(0,1){8.5}}
   \put(0,-30){\line(0,1){8.5}}
   \put(20,-30){\line(0,1){8.5}}
   \put(-20,30){\line(0,-1){8.5}}
   \put(0,30){\line(0,-1){8.5}}
   \put(20,30){\line(0,-1){8.5}}
   \put(-30,-20){\line(1,0){8.5}}
   \put(-30,0){\line(1,0){8.5}}
   \put(-30,20){\line(1,0){8.5}}
   \put(30,-20){\line(-1,0){8.5}}
   \put(30,0){\line(-1,0){8.5}}
   \put(30,20){\line(-1,0){8.5}}
   \put(-18.5,-20){\line(1,0){17}}
   \put(-18.5,0){\line(1,0){17}}
   \put(-18.5,20){\line(1,0){17}}
   \put(1.5,-20){\line(1,0){17}}
   \put(1.5,0){\line(1,0){17}}
   \put(1.5,20){\line(1,0){17}}
   \put(-20,-18.5){\line(0,1){17}}
   \put(0,-18.5){\line(0,1){17}}
   \put(20,-18.5){\line(0,1){17}}
   \put(-20,1.5){\line(0,1){17}}
   \put(0,1.5){\line(0,1){17}}
   \put(20,1.5){\line(0,1){17}}
   \put(20,1.5){\line(0,1){17}}
   \qbezier[100](1.06,21.06)(10,25)(18.93,21.06)
   \qbezier[100](-1.06,21.06)(-10,25)(-18.93,21.06)
   \qbezier[100](1.06,1.06)(10,5)(18.93,1.06)
   \qbezier[100](-1.06,1.06)(-10,5)(-18.93,1.06)
   \qbezier[100](1.06,-18.93)(10,-15)(18.93,-18.93)
   \qbezier[100](-1.06,-18.93)(-10,-15)(-18.93,-18.93)
   \qbezier[100](-18.93,-18.93)(-15,-10)(-18.93,-1.06)
   \qbezier[100](1.06,-18.93)(5,-10)(1.06,-1.06)
   \qbezier[100](21.06,-18.93)(25,-10)(21.06,-1.06)
   \qbezier[100](-18.93,1.06)(-15,10)(-18.93,18.93)
   \qbezier[100](1.06,1.06)(5,10)(1.06,18.93)
   \qbezier[100](21.06,1.06)(25,10)(21.06,18.93)
  \end{picture}
 \end{center}
 \vspace*{-1em}

 \caption{Two-dimensional lattice (with $L = 3$). The white circles are
 sites in $\Lambda_o$ and the black dots are sites in
 $\Lambda^\prime$. Electrons at a site can hop to another site if this
 site is connected to the original site with a line or a curve.}
 \label{fig:2-dim_lattice}
\end{figure}
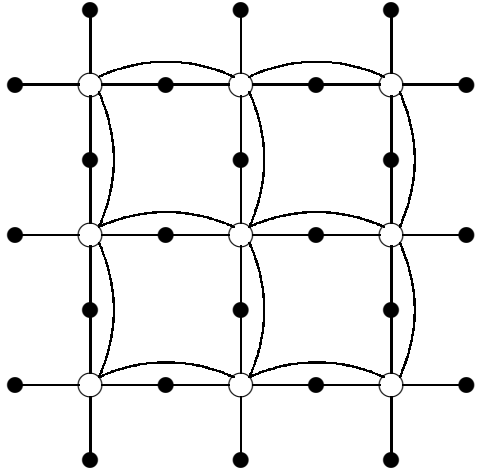

The lattice $\Lambda$ on which our deformed Hubbard model is defined is
decomposed into two sublattices
\begin{equation}
 \Lambda = \Lambda_o \cup \Lambda^\prime.
\end{equation}
$\Lambda_o$ is $d$-dimensional integer lattice with linear size $L$,
which is defined
\begin{equation}
 \Lambda_o :=
  \left\{
   x = (x_{1}, x_{2}, \cdots, x_{d}) \in {\mathbb Z}^{d}
   \biggl| |x_{j}| \leq \frac{L - 1}{2} \quad j = 1, 2, \cdots, d
  \right\}.
\end{equation}
$\Lambda^\prime$ can be further decomposed to
$\Lambda_{j}$ ($j=1, 2, \cdots, d$), {\it i.e.}
\begin{equation}
 \Lambda^\prime = \bigcup_{j = 1}^{d}\Lambda_{j}.
\end{equation}
$\Lambda_j$ is obtained as a half integer translation of
$\Lambda_o$ to $j$-th direction,
\begin{equation}
 \Lambda_{j} :=
  \left\{ x + e^{(j)} | x \in \Lambda_o \right\} \cup
  \left\{ x - e^{(j)} | x \in \Lambda_o \right\},
\end{equation}
where $e^{(j)}$ is defined
\begin{equation}
 e^{(j)} := (0, \cdots, 0,
  \begin{array}[t]{@{}l@{}}
   \frac{1}{2}, 0, \cdots, 0 ).\\
   \uparrow\\
   j\mbox{-th}
  \end{array}
\end{equation}
We show the lattice in the two-dimensional case in
Fig. \ref{fig:2-dim_lattice} as an example.

\subsection{Electron Operators and the Fock Space}

The creation and annihilation operators for an electron are denoted by
$c_{x, \sigma}^\dagger$ and $c_{x, \sigma}$. They obey the standard
anticommutation relations
\begin{equation}
 \{ c_{x, \sigma}, c_{y, \tau}^\dagger \}
  = \delta_{x, y} \delta_{\sigma, \tau}, ~~
  \{ c_{x, \sigma}, c_{y, \tau} \}
  = 0 = \{ c_{x, \sigma}^\dagger, c_{y, \tau}^\dagger \},
  \label{eq:anti-comm-rel_for_c}
\end{equation}
where $\{A, B\} = A B + B A$, for sites $x, y \in \Lambda$ and
spin coordinates $\sigma, \tau = \uparrow, \downarrow$.
We define no-electron state $\Phi_{\rm vac}$ by
\begin{equation}
 c_{x, \sigma} \Phi_{\rm vac} = 0
\end{equation}
for all $ x \in \Lambda$ and $\sigma = \uparrow, \downarrow$. We
construct a Fock space spanned by a basis
\begin{equation}
 \left\{
  \left( \prod_{x \in A} c_{x, \uparrow}^\dagger \right)
  \left( \prod_{x \in B} c_{x, \downarrow}^\dagger \right)
  \Phi_{\rm vac}
  \biggl| A, B \subset \Lambda
 \right\}.
\end{equation}
We also define a number operator $n_{x, \sigma}$
by $n_{x, \sigma} = c_{x, \sigma}^\dagger c_{x, \sigma}$ whose
eigenvalue represents a number of electrons at site $x$ with spin $\sigma$.
Note anticommutation relations
$\{ c_{x, \sigma}^\dagger, c_{x, \sigma}^\dagger \} = 0$ {\it i.e.}
$c_{x, \sigma}^\dagger c_{x, \sigma}^\dagger = 0.$
This relation implies the Pauli principle.
We employ the open boundary condition, when we
consider domain wall ground states.
This is realized by $c_{x, \sigma} = 0$ if
$|x_j| > L/2$ for some $j = 1, 2, \cdots , d$ with
$x = (x_{l})_{l = 1}^{d}$. We employ the periodic boundary condition,
when we consider the spin-wave excitation above the all-spin-up ground state.

\subsection{Deformed Flat-Band Hubbard Model}

Before we define the Hamiltonian, we introduce new operators
$\tilde{a}_{x, \sigma}^\dagger$ and $d_{x, \sigma}$ defined by
\begin{equation}
 \tilde{a}_{x, \sigma}^\dagger =
  \begin{cases}
   \displaystyle
   - q^{p(\sigma)/4} \sum_{j = 1}^{d}
   c_{x - e^{(j)}, \sigma}^\dagger
   + \lambda c_{x, \sigma}^\dagger
   - q^{-p(\sigma)/4} \sum_{j = 1}^{d}
   c_{x + e^{(j)}, \sigma}^\dagger
   & \mbox{if } x \in \Lambda_o
   \\
   \lambda^{-1} c_{x, \sigma}^\dagger
   & \mbox{if } x \in \Lambda^\prime
  \end{cases},
  \label{eq:def_of_a}
\end{equation}
and
\begin{equation}
 d_{x, \sigma} =
  \begin{cases}
   \lambda^{-1} c_{x, \sigma}
   & \mbox{if } x \in \Lambda_o
   \\
   \displaystyle
   q^{-p(\sigma)/4} c_{x - e^{(j)}, \sigma}
   + \lambda c_{x, \sigma}
   + q^{p(\sigma)/4} c_{x + e^{(j)}, \sigma}
   & \mbox{if } x \in \Lambda_{j}
  \end{cases},
  \label{eq:def_of_d}
\end{equation}
where $q$ is a complex parameter, $\lambda$ is a positive parameter and
$p(\sigma)$ takes $+1$ if $\sigma = \uparrow$ and $-1$ if $\sigma = \downarrow$.
And we formally define $\tilde{a}_{x, \sigma}^\dagger = 0$ and
$d_{x, \sigma} = 0$ if $|x_{j}| > L/2$
for some $j = 1, 2, \cdots , d$ with $x = (x_{l})_{l = 1}^{d}$.
This definitions correspond to the open boundary condition for the
original electron operators.
Note that these $\tilde{a}_{x, \sigma}^\dagger$ and $d_{x, \sigma}$ satisfy the
 anticommutation relations,
\begin{equation}
 \{\tilde{a}_{x, \sigma}^\dagger, d_{y, \tau} \} = \delta_{x, y}
  \delta_{\sigma, \tau}, \quad
  \{ \tilde{a}_{x, \sigma}^\dagger, \tilde{a}_{y, \tau} \} = 0 =
  \{ d_{x, \sigma}, d_{y, \tau} \}.
  \label{anticomm}
\end{equation}
We can easily obtain the following inverse relations of (\ref{eq:def_of_a}) and
(\ref{eq:def_of_d})
\begin{equation}
 c_{x, \sigma}^\dagger =
  \begin{cases}
   \displaystyle
   q^{p(\sigma)/4} \sum_{j = 1}^{d}
   \tilde{a}_{x - e^{(j)}, \sigma}^\dagger
   + \frac{1}{\lambda} \tilde{a}_{x, \sigma}^\dagger
   + q^{-p(\sigma)/4} \sum_{j = 1}^{d}
   \tilde{a}_{x + e^{(j)}, \sigma}^\dagger
   & \mbox{if } x \in \Lambda_o
   \\
   \lambda \tilde{a}_{x, \sigma}^\dagger & \mbox{if } x \in \Lambda^\prime
  \end{cases}
  ,
\end{equation}
and
\begin{equation}
 c_{x, \sigma} =
  \begin{cases}
   \lambda d_{x, \sigma} & \mbox{if } x \in \Lambda_o \\
   \displaystyle
   -q^{-p(\sigma)/4} d_{x - e^{(j)}, \sigma}
   + \frac{1}{\lambda} d_{x, \sigma}
   - q^{p(\sigma)/4} d_{x + e^{(j)}, \sigma}
   & \mbox{if } x \in \Lambda_{j}
  \end{cases}
  .
\end{equation}
The existence of inverse relations implies that the Fock space is also
spanned by another basis
\begin{equation}
 \left\{
  \left( \prod_{x \in A} \tilde{a}_{x, \uparrow}^\dagger \right)
  \left( \prod_{x \in B} \tilde{a}_{x, \downarrow}^\dagger \right)
  \Phi_{\rm vac}
  \biggl| A, B \subset \Lambda
 \right\}.
\end{equation}
This fact is useful to obtain the ground states.

The definition of our Hubbard Hamiltonian is given by
\begin{equation}
 H := H_{\rm hop} + H_{\rm int},
\label{Hamiltonian}
\end{equation}
where $H_{\rm hop}$ and $H_{\rm int}$ defined
\begin{equation}
 H_{\rm hop} = t \sum_{\sigma = \uparrow, \downarrow}
  \sum_{x \in \Lambda^\prime}
  d_{x, \sigma}^\dagger d_{x, \sigma}
\end{equation}
and
\begin{equation}
 H_{\rm int} = U \sum_{x \in \Lambda}
  n_{x, \uparrow} n_{x, \downarrow}
  \label{eq:H_int}
\end{equation}
with $t, U > 0$. The hopping Hamiltonian $H_{\rm hop}$ can be written
in the following form
\begin{equation}
 H_{\rm hop} = \sum_{x, y \in \Lambda} t_{x, y}^{(\sigma)}
  c_{x, \sigma}^\dagger c_{y, \sigma}
\end{equation}
where
\begin{equation}
 t_{x, y}^{(\sigma)} = ( t_{y, x}^{(\sigma)} )^{\ast}
  =
  \begin{cases}
   t d (|q|^{1/2} + |q|^{-1/2})
   & \mbox{if } x = y \in \Lambda_o\\
   t \lambda^{2}
   & \mbox{if } x = y \in \Lambda^\prime\\
   t \lambda q^{p(\sigma)/4}
   & \mbox{if } x \in \Lambda^\prime, y \in \Lambda_o
   \mbox{ with } [x] < [y] \mbox{ and } |x - y| = \frac{1}{2}\\
   t \lambda q^{-p(\sigma)/4}
   & \mbox{if } x \in \Lambda^\prime, y \in \Lambda_o
   \mbox{ with } [x] >[y] \mbox{ and } |x - y| = \frac{1}{2}\\
   t e^{-i p(\sigma) \theta/ 2}
   & \mbox{if } x, y \in \Lambda_o \mbox{ with } [x] >[y] \mbox{ and } |x-y|=1\\
   0 & \mbox{otherwise}
  \end{cases}
  .
\end{equation}
with a definition $[x]=\sum_{j=1} ^d x_j$. We parametrize%
\footnote{Throughout the present paper, we denotes a complex conjugate
of $\alpha \in {\mathbb C}$ by $\alpha^{\ast}$ and its absolute value
by $|\alpha|$. We also denote $|v|$ to represent a norm of a vector $v$
in $d$-dimensional Euclidean space and $|A|$ to represent the cardinality of
a set $A$.\label{fn1}}
$q = |q| e^{i \theta}$ by $0 \leq \theta < 2 \pi$%
.
Each term $t_{x, y}^{(\sigma)} c_{x, \sigma}^\dagger c_{y, \sigma}$
in the hopping Hamiltonian represents that an electron
with spin $\sigma$ hops from site $x$ to site $y$ with a probability
proportional to $|t_{x, y}^{(\sigma)}|^{2}$.

Since the interaction Hamiltonian $H_{\rm int}$ represents an on-site
repulsive interaction, this Hamiltonian can be regarded as a simplification of
the Coulomb interaction between two electrons.

Note that this system conserves the number of electron. The total
electron number operator $\hat{N}_{e}$ is defined by
\begin{equation}
 \hat{N}_{e} := \sum_{x \in \Lambda}
  \sum_{\sigma = \uparrow, \downarrow} n_{x, \sigma}.
\end{equation}
Since the Hamiltonian commutes with this operator, we can set the
electron number to an arbitrary filling. In the
present paper, we only consider that the
electron number is equal to $|\Lambda_o|$, namely,
the Hilbert space ${\cal H}$ is spanned by the following basis${}^{\ref{fn1}}$
\begin{equation}
 \left\{
  \left( \prod_{x \in A} c_{x, \uparrow}^\dagger \right)
  \left( \prod_{x \in B} c_{x, \downarrow}^\dagger \right)
  \Phi_{\rm vac}
  \biggl| A, B \subset \Lambda \mbox{ with }
  |A| + |B| = |\Lambda_o|
 \right\},
\end{equation}
or
\begin{equation}
 \left\{
  \left( \prod_{x \in A} \tilde{a}_{x, \uparrow}^\dagger \right)
  \left( \prod_{x \in B} \tilde{a}_{x, \downarrow}^\dagger \right)
  \Phi_{\rm vac}
  \biggl| A, B \subset \Lambda \mbox{ with }
  |A| + |B| = |\Lambda_o|
 \right\}.
\end{equation}

Let us discuss the symmetry of the model. First important symmetry is a
U(1) symmetry. We define spin operators at site $x$ by
\begin{equation}
 S_{x}^{(l)} := \sum_{\sigma, \tau = \uparrow, \downarrow}
  c_{x, \sigma}^\dagger
  \frac{{\cal P}_{\sigma, \tau}^{(l)}}{2} c_{x, \tau},
\end{equation}
where  ${\cal P}^{(l)}$ ($l = 1, 2, 3$) denote Pauli matrices
\begin{equation}
 {\cal P}^{(1)} =
  \begin{pmatrix}
   0 & 1 \\ 1 & 0
  \end{pmatrix}
  , \quad
  {\cal P}^{(2)} =
  \begin{pmatrix}
   0 & - i \\ i & 0
  \end{pmatrix}
  , \quad
  {\cal P}^{(3)} =
  \begin{pmatrix}
   1 & 0 \\ 0 & -1
  \end{pmatrix}
  .
\end{equation}
The Hamiltonian commutes with the third component of total spin operator
\begin{equation}
 [H, S_{\rm tot}^{(3)}] =
  H S_{\rm tot}^{(3)} - S_{\rm tot}^{(3)} H = 0,
\label{good}
\end{equation}
with
\begin{equation}
 S_{\rm tot}^{(l)} = \sum_{x \in \Lambda} S_{x}^{(l)}.
\end{equation}
We call the eigenvalue of $S_{\rm tot}^{(3)}$ magnetization.
We can classify energy eigenstate by the magnetization.
Note that this U(1) symmetry generated by $S_{\rm tot}^{(3)}$
is enhanced to an SU(2) symmetry in the case of
$q = 1$ {\it i.e.} Hamiltonian commutes with any component of total spin
operator.
In this case, this model becomes the original flat-band Hubbard
model given by Tasaki \cite{T0,T}.
Another important symmetry is generated by a product of parity and
spin rotation defined by
\begin{equation}
 \Pi = \Pi^{-1} = P \exp \left( i \pi S_{\rm tot}^{(1)} \right),
\end{equation}
where $P$ is a parity operator defined by
$P c_{x, \sigma} P = c_{-x, \sigma}$ and
$P c_{x, \sigma}^\dagger P = c_{-x, \sigma}^\dagger$.
$\Pi$ transforms $c_{x, \sigma}$ and $c_{x, \sigma}^\dagger$ to
$c_{-x, \overline{\sigma}}$ and $c_{-x, \overline{\sigma}}^\dagger$,
where $\overline{\sigma} = \downarrow$ if $\sigma = \uparrow$ or
$\overline{\sigma} = \uparrow$ if $\sigma = \downarrow$.
Note the following transformation of the total magnetization
$\Pi S_{\rm tot}^{(3)} \Pi = - S_{\rm tot}^{(3)}$. An energy eigenstate
with the total magnetization $M$ is transformed by $\Pi$ into another
eigenstate with the total magnetization $-M$, which belongs to the same
energy eigenvalue.
Note that the Hamiltonian of the XXZ model with a boundary field $h$
\begin{align}
 -J \sum_{x \in \Lambda_o} \sum_{j=1}^d
  \left[
   S_{x}^{(1)} S_{x+2e^{(j)}}^{(1)}
   + S_{x}^{(2)} S_{x+2e^{(j)}}^{(2)}
   + \frac{q + q^{-1}}{2} S_{x}^{(3)} S_{x+2e^{(j)}}^{(3)}
   +
   h S_{x}^{(3)} - h S_{x+2e^{(j)}}^{(3)} 
  \right],
\end{align}
has these two symmetries as well. Our deformation of the hopping Hamiltonian 
in the flat-band Hubbard model is one of the simplest way which preserves these 
two symmetries. If one does not want to add the XXZ Hamiltonian which leads to the 
ferromagnetism trivially, one reaches to our model naturally.

\section{Ground States\label{sec:GS}}

In this section, we obtain ground states of the model with the fixed
electron number $N_{e} = |\Lambda_o|$ on the basis of Tasaki's
construction method \cite{T0}.

\subsection{Construction of Ground States}

The representation of the hopping Hamiltonian in terms of
$d_{x, \sigma}$,
\begin{equation}
 H_{\rm hop} =
  t \sum_{\sigma = \uparrow, \downarrow} \sum_{x \in \Lambda^\prime}
  d_{x, \sigma}^\dagger d_{x, \sigma},
\end{equation}
indicates the positive semi-definiteness $H_{\rm hop} \geq 0$.
The positive semi-definiteness of the interaction Hamiltonian
$H_{\rm int} \geq 0$ is also clear because
$n_{x,\sigma} = c_{x, \sigma}^\dagger c_{x, \sigma} \geq 0$,
then the total Hamiltonian is also positive semi-definite
\begin{equation}
 H=H_{\rm hop}+H_{\rm int} \geq 0.
  \label{psd}
\end{equation}
First, we consider a fully polarized state $\Phi_\uparrow$ defined by
\begin{equation}
 \Phi_\uparrow =
  \left(
   \prod_{x \in \Lambda_o} \tilde{a}_{x, \uparrow}^\dagger
  \right)
  \Phi_{\rm vac}.
\end{equation}
We easily verify $H \Phi_\uparrow = 0$ from the anticommutativity
(\ref{anticomm}), and therefore $\Phi_\uparrow$ is a ground state
of $H$. Next, we determine all other ground states.

The conditions that a state $\Phi$ is a ground state are obviously
$H_{\rm hop} \Phi = 0$ and $H_{\rm int} \Phi = 0$. In other words,
\begin{equation}
 \tilde{a}_{x, \sigma} \Phi = 0
  \quad \mbox{for all} \quad x \in \Lambda^\prime
  \quad \mbox{with} \quad \sigma = \uparrow, \downarrow
\label{eq:cond-of-GS-1}
\end{equation}
and
\begin{equation}
 c_{y, \uparrow} c_{y, \downarrow} \Phi = 0
  \quad \mbox{for all} \quad y \in \Lambda.
  \label{eq:cond-of-GS-2}
\end{equation}
We expand $\Phi$ into the following series
\begin{equation}
 \Phi = \sum_{A, B}
  \psi(A, B)
  \left(
   \prod_{x \in A} \tilde{a}_{x, \uparrow}^\dagger
  \right)
  \left(
   \prod_{y \in B} \tilde{a}_{y, \downarrow}^\dagger
  \right)
  \Phi_{\rm vac} \label{eq:exp-of-GS-1},
\end{equation}
where the summation is taken over all $A, B \subset \Lambda$ with $|A|+|B|=|\Lambda_o|$.
The first condition (\ref{eq:cond-of-GS-1}) implies that $\psi(A, B)$
does not vanish only for $A, B \subset \Lambda_o$. The second condition
(\ref{eq:cond-of-GS-2}) for $y \in \Lambda_o$ means $\psi(A, B)$ takes 0
for $A \cap B \neq \emptyset$ with $A, B \subset \Lambda_o$.
Then we obtain the following form:
\begin{equation}
 \Phi =
  \sum_\sigma \phi(\sigma)
  \left(
   \prod_{x \in \Lambda_o} \tilde{a}_{x, \sigma_{x}}^\dagger
  \right) \Phi_{\rm vac}
  \label{eq:rep_of_GS_M}
\end{equation}
where the summation is taken over all possible spin configurations
$\sigma=(\sigma_{x})_{x \in \Lambda_o}$.
To satisfy the second condition (\ref{eq:cond-of-GS-2}) for
$y \in \Lambda_j$ ($j=1,2,\cdots, d$), the coefficient holds
\begin{equation}
 \phi(\sigma) = q^{\left[ p(\sigma_{y - e^{(j)}})
  - p(\sigma_{y + e^{(j)}}) \right]/2}
  \phi(\sigma_{y - e^{(j)}, y + e^{(j)}}),
  \label{eq:cond_for_phi}
\end{equation}
where $\sigma_{x, y}$ is spin configuration
obtained by the exchange $\sigma_{x}$ and $\sigma_{y}$ in the original
configuration $\sigma$.
This relation implies the uniqueness of the ground state with a fixed
total magnetization, since two arbitrary spin configurations with same
total magnetization can be
related by successive exchanges of two nearest neighbour spins.
Therefore the degeneracy of these ground states is exactly the same as
that in the SU(2) symmetric model. This degeneracy is also the same as
the ground states in the XXZ model \cite{ASW}.

Note that we can find the ``shift operator'' $S_q^-$ which makes the
ground state with magnetization $M$ from fully polarized ground state
$\Phi_\uparrow$ by acting certain times
\begin{equation}
 \Phi_M = (S_q^-)^{L^d-2M} \Phi_\uparrow,
\end{equation}
where $\Phi_M$ is the ground state with magnetization $M$. And $S_q^-$
can be written as
\begin{equation}
 S_{q}^{-} = \sum_{x \in \Lambda} q^{[x]}
  \tilde{a}_{x, \downarrow}^\dagger d_{x, \uparrow}.
\end{equation}

\subsection{Another Representation of Ground States}

To explore the nature of the ground state, we write the ground state in a
more explicit way as obtained by Gottstein and Werner in \cite{GW}.
We define the following electron operator creating a superposed state
\begin{equation}
\alpha_{x}^\dagger(\zeta) = \sum_{\sigma= \uparrow, \downarrow}
\eta_{x,\sigma} \tilde{a}_{x, \sigma}^\dagger,
\end{equation}
where we define a function of $x \in \Lambda$ and spin $\sigma$
\[
 \eta_{x, \sigma} = \zeta^{-p(\sigma)/2} q^{-p(\sigma)[x]/2}
\]
with $[x] = \sum_{j=1}^d x_j$.
We define a ground state $\Psi(\zeta)$ for an arbitrary complex number $\zeta$ by
\begin{equation}
 \Psi(\zeta) =
  \sum_{n = 0}^{L^d} \zeta^{n} (S_{q}^{-})^{n} \Phi_\uparrow
  =
  \left(
   \prod_{x \in \Lambda_{0}}
   \alpha_{x}^\dagger(\zeta)
  \right)
  \Phi_{\rm vac}.
  \label{eq:Psi(zeta)}
\end{equation}
One can see the localization property of the electrons
in this representation $\Psi(\zeta)$. 
The spin state of an electron at each site is completely determined. 
Unlike the ground state with a fixed 
total magnetization, one knows the conditional probability 
of the electron spin at a site $x$.
In principle, one can check whether this ground state is realized
or not by local observations. From this fact, the state 
$\Psi(\zeta)$ is expected to be healthy even in the
infinite-volume limit. We expect that the corresponding ground state 
to $\Psi(\zeta)$ is also a pure state in the infinite-volume limit,
as in the XXZ model.
Note that the expectation value of an arbitrary local operator
in the corresponding ground state $\Psi(\zeta)$ in the XXZ model 
is asymptotically equal to
that in $\Phi_M$ for $|\zeta|, |M| = O(1)$ 
\cite{BCNS}. 
We expect that the ground state $\Psi(\zeta)$ in our Hubbard model has 
many properties which are the same as the domain wall ground state in
XXZ model. In our Hubbard model, however, these are difficult to be
shown, since the state $\Psi(\zeta)$ defined here in the Hubbard model
is not a product state unlike the ground state in the XXZ model.

\subsection{Spin One-Point Functions}

Let us now consider expectation values of the spin operators in the
ground state $\Psi(\zeta)$. We denote an expectation value of an
operator $A$ in the ground state $\Psi(\zeta)$ by $\langle A \rangle_{\zeta}$.
The expectation value of a localized spin at site $x$ is written
\begin{align}
 \langle S_{x}^{(j)} \rangle_{\zeta}
 =
 \frac{1}{2}\sum_{\sigma, \tau = \uparrow, \downarrow}
 {\cal P}^{(j)}_{\sigma, \tau}
 \frac{(\Psi(\zeta), c_{x, \sigma}^\dagger c_{x, \tau} \Psi(\zeta))}
 {\| \Psi(\zeta) \|^{2}}
 =
 \frac{1}{2}\sum_{\sigma, \tau = \uparrow, \downarrow}
 {\cal P}^{(j)}_{\sigma, \tau}
 \frac{(c_{x, \sigma} \Psi(\zeta), c_{x, \tau} \Psi(\zeta))}
 {\| \Psi(\zeta) \|^{2}}.
\end{align}
The following anticommutation relations
\begin{equation}
 \{ c_{x, \sigma}, \alpha_{y}^\dagger(\zeta) \} =
\lambda \eta_{x, \sigma} \delta_{x, y},
  \label{eq:anti-com_for_c_alpha-o}
\end{equation}
for $x ,y \in \Lambda_o$, and
\begin{equation}
 \{ c_{x, \sigma}, \alpha_{y}^\dagger(\zeta)\}
  = - \eta_{x, \sigma}
  \left(
   q^{-1/4} \delta_{x - e^{(j)}, y}
   + q^{1/4} \delta_{x + e^{(j)}, y}
  \right),
  \label{eq:anti-com_for_c_alpha-prime}
\end{equation}
for $x \in \Lambda_{j}$ and $y \in \Lambda_o$ are useful to calculate the expectation
value.
These anticommutation relations (\ref{eq:anti-com_for_c_alpha-o}) and
(\ref{eq:anti-com_for_c_alpha-prime}) yield an equation
\begin{equation}
 c_{x, \sigma} \Psi(\zeta) = \eta_{x, \sigma}
  \Psi_{x}(\zeta).
\end{equation}
Here, the state $\Psi_{x}(\zeta)$ is defined by
\begin{equation}
  \Psi_{x}(\zeta) =
  \begin{cases}
   \displaystyle {\rm sgn}(x) \lambda
   \left(
   \prod_{y \neq x}
   \alpha_{y}^\dagger(\zeta)
   \right)
   \Phi_{\rm vac} & \mbox{if } x \in \Lambda_o \\
   \displaystyle
   \!\!\!
   \Biggl(
   {\rm sgn}(x - e^{(j)}) q^{-1/4}
   \!\!\!\!\!
   \prod_{y \neq x - e^{(j)}}
   \!\!\!\!
   \alpha_{y}^\dagger(\zeta)
   + 
   {\rm sgn}(x + e^{(j)}) q^{1/4}
   \!\!\!\!\!
   \prod_{y \neq x + e^{(j)}}
   \!\!\!\!
   \alpha_{y}^\dagger(\zeta)
   \Biggr)
   \Phi_{\rm vac}
   & \mbox{if } x \in \Lambda_{j}
  \end{cases}
  ,
\end{equation}
where ${\rm sgn}(x)$ takes $\pm 1$.
Then, the expectation value of $c_{x, \sigma}^\dagger c_{x, \tau}$ for
all $x \in \Lambda$ in the ground state $\Psi(\zeta)$ can be written in
terms of $\Psi_{x}(\zeta)$,
\begin{equation}
 \langle c_{x, \sigma}^\dagger c_{x, \tau} \rangle_{\zeta}
  =\eta_{x, \sigma}^{\ast} \eta_{x, \tau}
  \frac{\| \Psi_{x}(\zeta) \|^{2}}{\| \Psi(\zeta) \|^{2}}.
\end{equation}
Thus we obtain the representations of spin one-point functions at site
$x \in \Lambda$ in terms of an electron number density
$\langle n_{x} \rangle_{\zeta}$
($n_x := n_{x, \uparrow} + n_{x, \downarrow}$),
\begin{align}
 \langle S_{x}^{(1)} \rangle_{\zeta}
 = & \frac{\langle n_{x} \rangle_{\zeta}}{2}
 \frac{\zeta q^{[x]} + (\zeta q^{[x]})^{\ast}}{1 + |\zeta q^{x}|^{2}},
 \\
 \langle S_{x}^{(2)} \rangle_{\zeta}
 = & \frac{\langle n_{x} \rangle_{\zeta}}{2 i}
 \frac{\zeta q^{[x]} - (\zeta q^{[x]})^{\ast}}{1 + |\zeta q^{x}|^{2}},
 \\
 \langle S_{x}^{(3)} \rangle_{\zeta}
 = & \frac{\langle n_{x} \rangle_{\zeta}}{2}
 \frac{1 - |\zeta q^{[x]}|^{2}}{1 + |\zeta q^{[x]}|^{2}}.
\label{spin}
\end{align}
We expect that the electron number density in the ground state $\Psi(\zeta)$
is almost constant on $\Lambda_o$ or on $\Lambda^\prime$ respectively,
from the definition of $\Psi(\zeta)$. Indeed, in the one-dimensional model,
we can check this conjecture by the exact bounds
\cite{HI2}.

As in the domain wall ground state of the XXZ models discussed in
\cite{KN, M, KNS}, the two domains are distinguished by the sign of the
local order parameter $\langle S_{x} ^{(3)} \rangle_{\zeta}$. The domain
wall center is defined by zeros of
$\langle S_{x} ^{(3)} \rangle_{\zeta}$ which is located at $x$ with
$[x]= - \log_{|q|}|\zeta|$.
The function
$\frac{1}{2}\langle n_{x}\rangle_{\zeta} -|\langle S_{x}^{(3)} \rangle_{\zeta}|$
decays exponentially as $x$ is far away from the center.
This decay length defines the domain wall width
$1 / \log |q|$.
If the number density is almost constant on each sublattice
$\Lambda_o$ or $\Lambda^\prime$ as we conjectured, the behaviors of the
spin one-point functions are not controlled by the number density.
In large $\lambda$ limit for real $q > 1$, electrons are
completely localized at integer sites, and the spin one-point functions
are exactly the same as those obtained in the XXZ model defined on $\Lambda_o$.

For a complex $q=|q|e^{i \theta}$, one can see the spiral structure
with a pitch angle $\theta$. The vector
$\langle \vec{S}_x \rangle_{\zeta}
:= (\langle S_{x}^{(j)} \rangle_{\zeta})_{j = 1}^{3}$
is rotated with the angle $[x] \theta$ around the
third spin axis depending on the site $x$.
Note that this spiral structure of the ground state does not exist in
the XXZ model, though the complex anisotropy parameter $q=e^{i \theta}$
is possible in the XXZ Hamiltonian.
The corresponding model is described in the Tomonaga-Luttinger liquid
without ferromagnetic order in one dimension.

The translational symmetry in the infinite-volume limit is broken by the
domain wall or the spiral structure for finite $\log |\zeta|$.
Both symmetries generated by $S^{(3)}$ and $\Pi$ are broken
spontaneously as well.

\subsection{Spin Correlation Functions}

The spin correlation function can be also represented in terms of the
correlation function of the local electron number operators
\begin{align}
 \langle S_{x}^{(j)} S_{y}^{(l)} \rangle_{\zeta}
 = &
 \frac{1}{4} \sum_{\sigma, \tau, \sigma^\prime, \tau^\prime}
 \frac{
 \eta_{x, \sigma}^{\ast}
 {\cal P}_{\sigma, \tau}^{(j)} \eta_{x, \tau}
 }
 {|\zeta q^{[x]}|+|\zeta q^{[x]}|^{-1}}
 \frac{
 \eta_{y, \sigma^\prime}^{\ast}
 {\cal P}_{\sigma^\prime, \tau^\prime}^{(l)}
 \eta_{y, \tau^\prime}
 }{|\zeta q^{[x]}|+|\zeta q^{[x]}|^{-1}}
 \langle n_{x} n_{y} \rangle_{\zeta}.
\end{align}
We can rewrite
\begin{equation}
 \langle S_{x}^{(j)} S_{y}^{(l)} \rangle_{\zeta}
  = \langle S_{x}^{(j)} \rangle_{\zeta}
  \langle S_{y}^{(l)} \rangle_{\zeta}
  \frac{\langle n_{x} n_{y} \rangle_{\zeta}}
  {\langle n_{x} \rangle_{\zeta} \langle n_{y} \rangle_{\zeta}}.
\end{equation}
if $\lambda < \infty$.
If one estimates the correlation function of the local electron number
operators, one can check the cluster property of the ground state.
Actually this can be done for the one-dimensional model \cite{HI2}.

\section{Existence of Gapless Excitations\label{sec:Gapless}}

Here, we show an upper bound of excitation energy in $d \geq 2$
for sufficiently large finite volume under the open boundary condition. 
We generalize some parts of Matsui's argument for the product ground
state in the XXZ model in \cite{M} to those for the non-product ground
state in the flat-band Hubbard model. We estimate 
the energy in a trial state constructed by acting a local operator on 
a domain wall ground state $\Psi(\zeta)$.

\subsection{Low Energy Excitations}

Here, we show two results for low energy excitations in our model.

{\theorem (``Local gapless excitation'' above the ground state $\Psi(\zeta)$)
In the $d$-dimensional Hubbard model defined by the Hamiltonian
(\ref{Hamiltonian}) with $d \geq 2$ and the system volume
$|\Lambda_o|=L^d$, for an arbitrary $\zeta \in {\mathbb C}$ with
$|\log_{|q|}|\zeta|| < d(L-1)/2$ and an arbitrary $l$ with
$1<l \leq L+2 |\log_{|q|}|\zeta||/d$, there exists a local operator
$O_l$ defined on a compact support with a linear size $l$ and a constant
$F_1 > 0$ independent of the system size such that
\begin{equation}
 \frac{(O_l \Psi(\zeta), H O_l \Psi(\zeta))}{\| O_l \Psi(\zeta)\|^2} < F_1 U l^{-1},
  \label{eq:gapless2}
\end{equation}
and $(\Psi(\zeta),O_l \Psi(\zeta)) = 0$.
Moreover, there is an upper bound on the norm of a projected state $P_0 O_l \Psi(\zeta)$, 
where $P_0$ is the projection operator onto the space of ground states.
There exist constants
$L_1 > 0$, $R > 1$ and $F_2 > 0$ which are independent of the system size such that
\begin{equation}
 \frac{\| P_0 O_l \Psi(\zeta) \|^2}{\| O_l  \Psi(\zeta)\|^2} <
  F_2 l^{d+1} (L^d+1) R^{-2L^{d-1}}
  \mbox{ for } L > L_1.
  \label{eq:almost_unity}
\end{equation}
\label{th:gapless2}}

Here, we describe some physical meanings of Theorem \ref{th:gapless2}. 
We emphasize that the excited state in Theorem \ref{th:gapless2} is
constructed by acting a local operator which consists of finite number of
electron operators $c_{x,\sigma}^\dag$ and $c_{x,\sigma}$. 
As discussed when we defined the
ground state $\Psi(\zeta)$, one can confirm whether the system takes the
ground state $\Psi(\zeta)$ or not by local observations.
After one checks the ground state $\Psi(\zeta)$ once, one can
obtain a locally deformed state, say $O_l \Psi(\zeta)$, by a local
operation to the system. In this sense, Theorem \ref{th:gapless2} claims
that one can change the state of the system from the ground state
$\Psi(\zeta)$ by the local operation with energy as small as one wants. 
Particularly, the second result in Theorem \ref{th:gapless2}
guarantees that the deformed state $O_l \Psi(\zeta)$ has a non-zero
orthogonal component to all ground states of the model. This fact
implies that the local operation represented in $O_l$ is really
effective to deform the ground state $\Psi(\zeta)$ of the
system. Therefore, we can claim that there exists a gapless excitation
above the domain wall ground state $\Psi(\zeta)$.

\paragraph{Remark}
Theorem \ref{th:gapless2} should imply the existence of a gapless
excitation in the infinite-volume limit under the condition of the fixed
electron number, if the corresponding ground state in the infinite
system to $\Psi(\zeta)$ were shown to be pure and unique as in the XXZ
model \cite{M}.
\vspace*{1em}

We can prove the following property of the lowest excitation energy
eigenvalue of $H$ directly from Theorem \ref{th:gapless2}.

{\corollary (Spectra of a finite system with open boundary condition)
Suppose the $d$-dimensional Hubbard model defined by the Hamiltonian
(\ref{Hamiltonian}) with $d \geq 2$. Let $E_1$ be the lowest energy
eigenvalue of excitation in the model with the system volume
$|\Lambda_o|=L^d$ under the open boundary condition. There exist
constants $L_2 >0$ and $F_3 > 0$ which are independent of the system
size such that
\begin{equation}
 E_1 \leq F_3 U L^{-1} \quad \mbox{for} \quad L \geq L_2.
  \label{eq:gapless1}
\end{equation}
\label{th:gapless1}}

\subsection{Expansion in the Original Basis}
Here we prepare for the proof of Theorem \ref{th:gapless1} and \ref{th:gapless2}. 
To evaluate the inner product between two states, we represent them in
terms of original electron operator $c_{x, \sigma}^\dagger$. This
representation has good properties which help us to estimate expectation
values of observables.

\subsubsection{Space of Configurations}

To introduce a representation of states in terms of original electron
operators, we define the decoration of a site $x \in \Lambda$ by a set
$$
\bar{x} :=\bigcup_{j=1} ^d \{x,	 x + e^{(j)},  x - e^{(j)} \} \cap \Lambda.
$$
Note $\Lambda = \cup_{x \in \Lambda_o}\bar{x}$, which is not a disjoint union.
Also we define the decoration $\bar{X}$ of a subset $X \subset \Lambda_o$ by
$$
\bar{X}:= \bigcup_{x \in X}\bar{x}.
$$
To expand the ground state $\Psi(\zeta)$ in terms of an orthogonal basis, we
define the function $\xi_{x,\sigma}(\zeta)$ for $x \in \Lambda$ by
\begin{equation}
 \xi_{x,\sigma}(\zeta)	=
  \begin{cases}
   \displaystyle \lambda \eta_{x, \sigma} = \lambda \zeta^{-p(\sigma)/2}
q^{- p(\sigma) [x]/2}
    & \mbox{for } x \in \Lambda_o \\
   \displaystyle
   - \eta_{x, \sigma}=	-\zeta^{- p(\sigma)/2}
q^{- p(\sigma) [x]/2}
  & \mbox{for } x \in \Lambda'
  \end{cases}
  .
\label{amplitude}
\end{equation}
Since the operator $\alpha_x^\dag(\zeta)$ is written in terms of this function
$$
\alpha_x^\dag(\zeta) =	\sum_{y \in \bar{x}} \sum_{\sigma=\uparrow, \downarrow}
\xi_{y,\sigma}(\zeta) c_{y,\sigma} ^\dag,
$$
the ground state is represented in
\begin{equation}
\Psi(\zeta) = \left(\prod_{x \in \Lambda_o} \sum_{y \in \bar{x}} \sum_{\sigma=\uparrow, \downarrow}
\xi_{y,\sigma}(\zeta) c_{y,\sigma} ^\dag \right)\Phi_{\rm vac},
\label{gsc}
\end{equation}
To represent the ground state in terms of the electron creation operators on sites,
we define a set $\mathcal{C}_{\rm R}$ of all configurations for the ground state.
We define a position configuration for the ground state by a one-to-one mapping
$f: \Lambda_o \rightarrow \Lambda $ with a constraint $f(x) \in \bar{x}$
for each $x \in \Lambda_o$.  This mapping $f$ selects a site in each
decoration $\bar{x}$. We denote a set of
all position configurations for the ground state by $\mathcal{P}_{\rm R}$.
Also we define a spin configuration for the ground state
by a mapping $g: \Lambda_o \rightarrow \{\uparrow, \downarrow \}$.
We denote a set of all spin configurations for the ground state by $\mathcal{S}$.
A position configuration $f \in \mathcal{P}_{\rm R}$ and a spin configuration $g \in
\mathcal{S}$ define
a configuration in $\mathcal{C}_{\rm R}$
which is a mapping $\varphi: \Lambda_o \rightarrow \Lambda \times \{\uparrow, \downarrow \}$
such that $\varphi:x  \longmapsto \varphi (x)=(f(x), g(x))$ for $x \in \Lambda_o$.
An arbitrary configuration is also a one-to-one mapping.
Then the ground state (\ref{gsc}) is represented as a summation over all configurations
in orthogonal basis
\begin{equation}
\Psi(\zeta) = \sum_{\varphi \in \mathcal{C}_{\rm R}}
\left(\prod_{x \in \Lambda_o} \xi_{\varphi (x)}(\zeta)
c_{\varphi (x)} ^\dag \right)\Phi_{\rm vac}.
\label{configuration}
\end{equation}
Several terms in this summation over all configurations
are linearly dependent, and they cancel each other.
Therefore the summation over all configurations in $\mathcal{C}_{\rm R}$
is reducible.

\subsubsection{Simple Loop}

To obtain an irreducible configuration space, we consider two different
configurations $\varphi$ and $\varphi'$ in $\mathcal{C}_{\rm R}$.
Assume that two terms defined by $\varphi$ and $\varphi'$ are linearly
dependent, namely, there exists a number $C$,
\begin{equation}
\left(\prod_{x \in \Lambda_o}
\xi_{\varphi(x)}(\zeta)c_{
\varphi(x)} ^\dag \right)\Phi_{\rm vac}
=C \left(\prod_{x \in \Lambda_o}
\xi_{ \varphi'(x)}(\zeta)c_{
\varphi'(x)} ^\dag \right)\Phi_{\rm vac}.
\label{ld}
\end{equation}
We say that two configurations $\varphi, \varphi' \in \mathcal{C}_{\rm R}$
are linearly dependent if the terms defined by $\varphi$ and
$\varphi'$ are linearly dependent.
This relation implies
\begin{equation}
\{ \varphi(x) | x \in \Lambda_o \} = \{ \varphi'(x) | x \in \Lambda_o \},
\label{configid}
\end{equation}
otherwise one cannot obtain the relation (\ref{ld}) for any number $C$.
Nonetheless, $\varphi \neq \varphi'$ implies $\varphi(x) \neq \varphi'(x)$
for some $x \in \Lambda_o$.  Let us consider a set of sites for two
linearly dependent $\varphi, \varphi' \in \mathcal{C}_{\rm R}$
\begin{equation}
X(\varphi,\varphi')=\{x \in \Lambda_o | \varphi(x) \neq \varphi'(x)  \}.
\label{X}
\end{equation}
To study properties of this set, we define several terms.
We say that a sequence 
$$
\{x_{m+1}, x_{m+2}, \cdots, x_{m+n} \} \subset \Lambda_o
$$
for arbitrary positive even integers $m$ and $n$ is a simple loop with a length $n$,
if 
$\bar{x}_k \cap \bar{x}_{k+1} \neq \emptyset$ for $k=m+1, \cdots, m+n-1$
and $\bar{x}_{m+n} \cap \bar{x}_{m+1} \neq \emptyset$.
We say that a configuration $\varphi=(f, g) \in \mathcal{C}_{\rm R}$
for the ground state has a simple loop $\{x_{m+1}, x_{m+2}, x_{m+3}, \cdots, x_{m+n} \}$,
if $f(x_k) \in \bar{x}_{k+1}$  for $k=m+1, m+2, \cdots,  m+n-1$
and $f(x_{m+n}) \in \bar{x}_{m+1}$. 

\subsubsection{Loop Decomposition of Linearly Dependent Configurations}

Here, we show the following lemma.
{\lemma \label{lemma:3}
Let $n$ be a number of elements of the 
set $X(\varphi,\varphi')$ given in {\rm(\ref{X})} for two linearly dependent configurations
$\varphi, \varphi' \in \mathcal{C}_{\rm R}$. 
There exist a positive integer $N$ and some positive even integers
$0=m_0 < m_1 < \cdots < m_N = n$ for $X(\varphi,\varphi')$,
such that $X(\varphi,\varphi')$ can be written in a disjoint union of some simple loops
$$
X(\varphi,\varphi')= \bigcup_{j=1} ^N 
\{x_{m_{j-1}+1}, x_{m_{j-1}+2}, x_{m_{j-1}+3}, \cdots, x_{m_j} \}
$$
and
$
\varphi(x_{k-1}) = \varphi'(x_{k})$ for   $  m_{j-1}+1 < k \leq m_j $
and $  \varphi(x_{m_j}) = \varphi'(x_{m_{j-1}+1}).
$ 
Therefore, both configurations $\varphi$ and  $\varphi'$
have each simple loop
.}

\paragraph{Proof}
We attach indices to all sites in $X(\varphi,\varphi')$
in the following inductive manner. Let $x_1$ be an arbitrary site
in $X(\varphi,\varphi')$.
From the relation (\ref{configid}),
there exists a site $x_2 \in X(\varphi,\varphi')$ for the site $x_1$
such that $\varphi(x_1)=\varphi'(x_2).$	 Note $x_1 \neq x_2$
and $\bar{x}_1 \cap \bar{x}_2 \neq \emptyset$. By the relation
$\varphi(x_2) \neq \varphi'(x_2)$ there exists
$x_3 \in  X(\varphi,\varphi')$ for $x_2$ such that
$\varphi(x_2)=\varphi'(x_3)$
and $x_3 \neq x_1, x_2.$ There exists
$x_4 \in  X(\varphi,\varphi')$ for $x_3$ such that
$\varphi(x_3)=\varphi'(x_4)$ and $x_4 \neq x_1, x_2, x_3.$
If $\varphi(x_4)=\varphi'(x_1)$, then we obtain a simple loop
$\{ x_1, x_2, x_3, x_4 \}$ with a length $m_1=4$
and both configurations $\varphi$
and $ \varphi'$ have this simple loop.
This is  because $\bar{x}_4 \cap \bar{x}_1 \neq \emptyset$. If 
$\varphi(x_{4}) \neq \varphi'(x_1)$,
we define $x_5 \in X(\varphi,\varphi')$
such that $\varphi(x_4)=\varphi'(x_5)$ and $x_5 \neq x_1, x_2, x_3, x_4$.
We assume already defined
$x_{k-1} \in X(\varphi,\varphi')$  for an arbitrary natural number $k \leq n$
with $x_{k-1} \neq x_i$ for any $i = 1, 2, \cdots, k-2$, such that
$\varphi(x_{j})=\varphi'(x_{j+1})$ for $j=1, \cdots, k-2$. If $\varphi(x_{k-1}) =
\varphi'(x_1)$, then we obtain a simple loop
$\{x_1, x_2, x_3, \cdots, x_{k-1} \}$ with a length $m_1=k-1$
and both configurations $\varphi$
and $ \varphi'$ have this loop.
If $\varphi(x_{k-1}) \neq \varphi'(x_1)$,
we define $x_{k} \in X(\varphi,\varphi')$
such that $\varphi(x_{k-1})=\varphi'(x_{k})$ and $x_{k} \neq x_{i}$
for any $i=1, \cdots, k-1$. This can be proved as follows.
If $x_k=x_i$ for some $i =2, 3, \cdots, k-1$, then $\varphi(x_{k-1})=\varphi'(x_k)
=\varphi'(x_i)=\varphi(x_{i-1})$.
This equality and the definition of a one-to-one mapping
imply $x_{k-1} = x_{i-1}$, which
contradicts the assumption of the inductivity.
Also the assumption $\varphi(x_{k-1}) \neq \varphi'(x_1)$
excludes $x_k=x_1$. Thus, $x_k$ cannot be any element in $\{ x_1, x_2, x_3, \cdots, x_{k-1} \}$.
Note $\bar{x}_{k-1} \cap \bar{x}_{k} \neq \emptyset$.
There exists a number $m_1 \leq n$ such that
$\varphi(x_{m_1}) = \varphi'(x_1)$ and
$\bar{x}_{m_1} \cap \bar{x}_{1} \neq \emptyset$.
We obtain a simple loop $\{x_1, x_2, x_3, \cdots, x_{m_1} \}$
with a length $m_1$, and 
both configurations $\varphi$ and $\varphi'$ have this simple loop. 
If $X(\varphi,\varphi') \setminus  \{x_1, x_2, x_3, \cdots, x_{m_1} \}= \emptyset$, then $X(\varphi,\varphi')=
\{x_1, x_2, x_3, \cdots, x_{m_1} \}$ is a simple loop itself with
a length $n=m_1$. If
$X(\varphi,\varphi') \setminus \{x_1, x_2, x_3, \cdots, x_{m_1} \} \neq \emptyset$,
then $n-m_1 > 0$ and we continue to attach
indices $m_1+1, \cdots, n$ to the elements
in $X(\varphi,\varphi') \setminus
\{x_1, x_2, x_3, \cdots, x_{m_1} \}$. By attaching the indices
$m_1+1, \cdots, n$, we can continue to identify a subset of
$X(\varphi,\varphi') \setminus
\{x_1, x_2, x_3, \cdots, x_{m_1} \}$ to a simple loop that both configurations have.
Finally, we can write the set $X(\varphi,\varphi')$
as a disjoint union of some simple loops by attaching
the indices $1, 2, 3, \cdots, n$ to all the
elements in $X(\varphi,\varphi')$. 
Both configurations $\varphi$ and $\varphi'$ have each simple loop  
$\{x_{m_{j-1}+1}, x_{m_{j-1}+2}, x_{m_{j-1}+3}, \cdots, x_{m_j} \}$ with a length $m_j-m_{j-1}$, 
and  for two sites in each loop we have 
$
\varphi(x_{k-1}) = \varphi'(x_{k})
$ 
for $m_{j-1}+1 < k \leq m_j$ and 
$\varphi(x_{m_j}) = \varphi'(x_{m_{j-1}+1})$.
Thus, we have proved the lemma.
\qed

\subsubsection{Irreducible Configuration Space}

We define the irreducible set of all configurations
for the ground state by
$$
\mathcal{C}:= \{ \varphi \in \mathcal{C}_{\rm R} \ | \ 
\varphi \ \mbox{has no simple loop} \}.
$$
Here, we obtain the following lemma by proving that the terms given by
two configurations with a common simple loop are cancelled.
{\lemma \label{lemma:4} 
The ground state is represented in the summation over
irreducible set of all configurations 
\begin{equation}
\Psi(\zeta) = \sum_{\varphi \in \mathcal{C}}
\left(\prod_{x \in \Lambda_o} \xi_{\varphi (x)}(\zeta)
c_{\varphi (x)} ^\dag \right)\Phi_{\rm vac}.
\label{irrconfiguration}
\end{equation}
}

\paragraph{Proof}
From Lemma \ref{lemma:3},
we consider a configuration $\varphi \in \mathcal{C}_{\rm R}$
which has a simple loop $\{x_1, \cdots, x_n \}$
with the length $n=2m$. We show that this
configuration
$\varphi$ has a counter contribution which cancels
the contribution from $\varphi$.
For the configuration $\varphi$, there exists a unique
configuration $\varphi'$
with the same simple loop such that
$X(\varphi,\varphi')=\{x_1, \cdots, x_n \}$.
In this case, $\varphi(x_k)=\varphi'(x_{k+1})$
for any natural number $k \leq n-1$
and $\varphi(x_n)=\varphi'({x}_{1})$. Then, the
fermion statistics of the electron operators
gives the following relation
$$
\prod_{k=1} ^{2m} c_{\varphi(x_k)} ^\dag
= \prod_{k=1} ^{2m} c_{\varphi'(x_{k+1})} ^\dag= -
\prod_{k=1} ^{2m} c_{\varphi'(x_{k})} ^\dag,
$$
where we define $x_{2m+1} =x_1$. This relation implies $C=-1$
in the relation (\ref{ld}),
therefore the contributions of $\varphi$ and $\varphi'$
in (\ref{configuration}) cancel each other.
Now, we rewrite the representation (\ref{configuration})
of the ground state into summation over independent terms.
As discussed above, all configurations which
give linearly dependent terms in the representation
(\ref{configuration}) have at least one simple loop and all terms are cancelled.
Therefore, the representation
(\ref{configuration}) of ground state can be rewritten into a
summation over all configurations with no simple loop,
which consists of only linearly independent terms.
\qed \vspace*{1em}

The set $\mathcal{C}$ of all configurations can be decomposed into
a set $\mathcal{P}$ of all position configurations and
a set $\mathcal{S}$ of all spin configurations, where the set of all position configurations
is defined by
$$
\mathcal{P}:=\{f \in \mathcal{P}_{\rm R} \ | \
f \ \mbox{has no simple loop} \},
$$
as well as $\mathcal{C}$. On the other hand, all spin configurations have no constraint.
Therefore, after summing over all the spin configurations,
the ground state is represented in the summation over
irreducible set of all position configurations 
\begin{equation}
\Psi(\zeta) = \sum_{f \in \mathcal{P}}
\left[\prod_{x \in \Lambda_o}( \xi_{f(x),\uparrow}(\zeta)
c_{f(x),\uparrow} ^\dag +\xi_{f(x),\downarrow}(\zeta)
c_{f(x),\downarrow} ^\dag) \right]\Phi_{\rm vac}.
\label{irrconfiguration2}
\end{equation}

\subsection{Proof of Theorem \ref{th:gapless2}}

We define the following hyper cube on the lattice with a positive number
$l$
\begin{align}
Y_l :=
 \Biggl(
 \Biggl\{ (x_1, x_2, \cdots, x_d) \in {\mathbb R}^d \
 \Big| 
 \bigl|[x]\bigr|
 \leq \frac{l-1}{2}, & \ | x_j| < \frac{l-1}{2} 
 \mbox{ for all } j=1,2, \dots, d 
 \Biggr\} +
 \nonumber \\
 & \frac{z}{\sqrt{d|z|}}(1,1,\cdots,1)
 \Biggr) \cap \Lambda_o
 ,
\end{align}
where we define $z\in{\mathbb R}$ by $|\zeta| = q^{-z}$.
Note that the sublattice $Y_l$ has a linear size $l$.
We define a local operator
\begin{equation}
\tilde{S}^{(3)}_{Y_l} = \frac{1}{2}
 \sum_{x \in Y_l} \sum_{\sigma, \tau = \uparrow, \downarrow}
 \tilde{a}_{x, \sigma}^\dagger {\cal P}^{(3)}_{\sigma, \tau}  d_{x, \tau},
\end{equation}
and its deviation from the ground state expectation value
\begin{equation}
\delta \tilde{S}^{(3)}_{Y_l} = \tilde{S}_{Y_l}^{(3)} -
 \frac{(\Psi(\zeta), \tilde{S}_{Y_l}^{(3)} \Psi(\zeta))}{\| \Psi(\zeta) \|^2}
 .
\end{equation}
We define a normalized state $\tilde{\Psi}(\zeta)$ by
\begin{equation}
 \tilde{\Psi}_l(\zeta) := 
\frac{\delta \tilde{S}_{Y_l}^{(3)} \Psi(\zeta)}{\|\delta \tilde{S}_{Y_l}^{(3)} \Psi(\zeta)\|},
\label{normalized} 
\end{equation}
which is obviously orthogonal to the ground state $\Psi(\zeta)$.

First, we evaluate the norm of $\delta \tilde{S}_{Y_l}^{(3)} \Psi(\zeta)$.
\begin{equation}
\| \delta \tilde{S}_{Y_l}^{(3)} \Psi(\zeta) \|^2=	 \|(\tilde{S}^{(3)}_{Y_l} -
\langle \tilde{S}^{(3)}_{Y_l} \rangle_\zeta) \Psi(\zeta)\|^2.
\end{equation}
Lemma \ref{lemma:4} allows us to represent the state in a summation over
the configurations. If we define an indicator function $\chi$ by
$\chi[{\rm true}]=1$ and $\chi[{\rm false}]=0$, the state can be represented in
\begin{align}
 \tilde{S}_{Y_l}^{(3)} \Psi(\zeta) =
 \sum_{y \in Y_l}
 \displaystyle \sum_{f \in {\cal P}}
 \chi[f(y) = y] ~ {\rm sgn}(y)
 \sum_{w \in \overline{y}}
 (
 \xi_{w, \uparrow}(\zeta) c_{w, \uparrow}^\dagger -
 \xi_{w, \downarrow}(\zeta) c_{w, \downarrow}^\dagger
 ) \times
 \nonumber \\
 \left[
 \prod_{x \in \Lambda_o \backslash \{ y\}}
 (
 \xi_{f(x),\uparrow}(\zeta) c_{f(x),\uparrow}^\dag +
 \xi_{f(x),\downarrow}(\zeta) c_{f(x),\downarrow}^\dag
 )
 \right] \Phi_{\rm vac},
\end{align}
where ${\rm sgn}(y)=\pm 1$ is a sign factor coming from the fermion
statistics. Note that for an arbitrary $f\in {\cal P}$ with
$f(y)\neq y$, there exists $g \in {\cal P}$ such that $g(y)=y$ and
$f|_{\Lambda_o\backslash\{y\}}=g|_{\Lambda_o\backslash\{y\}}$.
Thus we can represent $\tilde{S}_{Y_l}^{(3)} \Psi(\zeta)$ by
\begin{align}
 \tilde{S}_{Y_l}^{(3)} \Psi(\zeta)
 = \sum_{y \in Y_l}\sum_{f \in \mathcal{P}}
 \left[
 \prod_{x \in \Lambda_o}( \xi_{f(x),\uparrow}(\zeta)
 c_{f(x),\uparrow} ^\dag
 +
 (-1)^{\chi[x=y]}\xi_{f(x),\downarrow}(\zeta)
 c_{f(x),\downarrow} ^\dag)
 \right]\Phi_{\rm vac} + \Phi_{\perp}.
 \label{irrconfiguration3}
\end{align}
The residual state $\Phi_{\perp}$ is orthogonal to the first term in the
right hand side as well as any ground state $\Psi(\zeta)$, since
$\Phi_\perp$ has no term written in a basis of 
irreducible configurations ${\cal C}$.
The ground state expectation value is represented as
\begin{equation}
\langle \tilde{S}^{(3)}_{Y_l} \rangle_\zeta = \frac{1}{2\|\Psi(\zeta)\|^2}\sum_{f \in \mathcal{P}}
\sum_{x_1 \in Y_l}
\frac{|\xi_{f(x_1),\uparrow}|^2-
|\xi_{f(x_1),\downarrow}|^2}{|\xi_{f(x_1),\uparrow}|^2+|\xi_{f(x_1),\downarrow}|^2}
\prod_{x \in \Lambda_o}(|\xi_{f(x),\uparrow}|^2+|\xi_{f(x),\downarrow}|^2),
\end{equation}
where $\mathcal{P}$ denotes the irreducible set of all the position configurations
of the ground state.
Another important part in the norm of $\delta \tilde{S}_{Y_l}^{(3)} \Psi(\zeta)$ is
\begin{align}
 \| \tilde{S}_{Y_l}^{(3)} \Psi(\zeta) \|^2
 =
 \sum_{f \in \mathcal{P}}
 \frac{1}{4}
 \left\{
 \sum_{x_1 =x_2 \in Y_l} 1
 +
 \sum_{x_1 \neq x_2 \in Y_l}
 \frac{|\xi_{f(x_1),\uparrow}|^2-|\xi_{f(x_1),\downarrow}|^2}
 {|\xi_{f(x_1),\uparrow}|^2+|\xi_{f(x_1),\downarrow}|^2}
 \frac{|\xi_{f(x_2),\uparrow}|^2-|\xi_{f(x_2),\downarrow}|^2}
 {|\xi_{f(x_2),\uparrow}|^2+|\xi_{f(x_2),\downarrow}|^2}\right\}
 \times
 \nonumber \\
 \prod_{x \in \Lambda_o}
 (|\xi_{f(x),\uparrow}|^2+|\xi_{f(x),\downarrow}|^2)+ \|\Phi_{\perp} \|^2
 \nonumber \\
 \geq
 \sum_{f \in \mathcal{P}} \frac{1}{4}
 \left\{
 \sum_{x_1 \in Y_l}
 \left[
 1 -
 \left(
 \frac{|\xi_{f(x_1),\uparrow}|^2-|\xi_{f(x_1),\downarrow}|^2}
 {|\xi_{f(x_1),\uparrow}|^2+|\xi_{f(x_1),\downarrow}|^2}
 \right)^2
 \right]
 + 
 \left( \sum_{x_1 \in Y_l}
 \frac{|\xi_{f(x_1),\uparrow}|^2-|\xi_{f(x_1),\downarrow}|^2}
 {|\xi_{f(x_1),\uparrow}|^2+|\xi_{f(x_1),\downarrow}|^2}
 \right)^2
 \right\}
 \times
 \nonumber \\
 \prod_{x \in \Lambda_o}
 (|\xi_{f(x),\uparrow}|^2+|\xi_{f(x),\downarrow}|^2).
 \nonumber
\end{align}
Thus, the norm of $\delta \tilde{S}_{Y_l}^{(3)} \Psi(\zeta)$ has the
following lower bound
\begin{align}
 & \| \delta \tilde{S}_{Y_l}^{(3)} \Psi(\zeta) \|^2
 \geq
 \sum_{f \in \mathcal{P}}
 \Biggl\{
 \sum_{x_1 \in Y_l} \frac{1}{4}
 \left[
 1 -
 \left(
 \frac{|\xi_{f(x_1),\uparrow}|^2-|\xi_{f(x_1),\downarrow}|^2}
 {|\xi_{f(x_1),\uparrow}|^2+|\xi_{f(x_1),\downarrow}|^2}
 \right)^2
 \right]
 \nonumber \\
 & \quad +
\frac{1}{4} \left[
 \sum_{x_1 \in Y_l} 
 \frac{|\xi_{f(x_1),\uparrow}|^2-|\xi_{f(x_1),\downarrow}|^2}{|\xi_{f(x_1),\uparrow}|^2+|\xi_{f(x_1),\downarrow}|^2}
 -2 \langle S^{(3)}_{Y_l} \rangle_\zeta  \right]^2
 \Biggr\}
 \prod_{x \in
 \Lambda_o}(|\xi_{f(x),\uparrow}|^2+|\xi_{f(x),\downarrow}|^2)
 \nonumber \\
 & \geq
 \sum_{f \in \mathcal{P}} \sum_{x_1 \in Y_l} \frac{1}{4}
 \left[1 - \left(\frac{|\xi_{f(x_1),\uparrow}|^2-|\xi_{f(x_1),\downarrow}|^2}
 {|\xi_{f(x_1),\uparrow}|^2+|\xi_{f(x_1),\downarrow}|^2} \right)^2 \right]
 \prod_{x \in \Lambda_o}
 (|\xi_{f(x),\uparrow}|^2+|\xi_{f(x),\downarrow}|^2)
 \nonumber \\
 & \geq \sum_{x_1 \in Y_l} \frac{1}{4}
 \left[
 1 -
 \left(
 \frac{|q|^{|[x_1]-z|+1/2}-|q|^{-|[x_1]-z|-1/2}}
 {|q|^{|[x_1]-z|+1/2}+
 |q|^{-|[x_1]-z|-1/2}} \right)^2 \right]
 \sum_{f \in \mathcal{P}}
 \prod_{x \in \Lambda_o}
 (|\xi_{f(x),\uparrow}|^2+|\xi_{f(x),\downarrow}|^2).
\end{align}
If we define $G_1 = (|q|^\frac{1}{2}+|q|^{-\frac{1}{2}})^{-2}$, we obtain
\begin{equation}
 \|\delta \tilde{S}_{Y_l}^{(3)} \Psi(\zeta)\|^2
  \geq
  G_1 l^{d-1} \|\Psi(\zeta)\|^2.
  \label{eq:norm}
\end{equation}

Next, we estimate $(\delta \tilde{S}_{Y_l}^{(3)} \Psi(\zeta), H \delta \tilde{S}_{Y_l}^{(3)} \Psi(\zeta))$.
The hopping energy of $\delta \tilde{S}_{Y_l}^{(3)} \Psi(\zeta)$ vanishes
\begin{equation}
 (\delta \tilde{S}_{Y_l}^{(3)} \Psi(\zeta), H_{\rm hop} \delta \tilde{S}_{Y_l}^{(3)} \Psi(\zeta))= 0,
\end{equation}
since $\delta \tilde{S}_{Y_l}^{(3)} \Psi(\zeta)$ consists of only operators
$a_{x, \sigma}^\dag$ with $x \in \Lambda_o$ acting on $\Phi_{\rm vac}$.
The interaction term
\begin{align}
 (\delta \tilde{S}_{Y_l}^{(3)} \Psi(\zeta),
 H_{\rm int} \delta \tilde{S}_{Y_l}^{(3)} \Psi(\zeta))
 =
 (\tilde{S}^{(3)}_{Y_l}\Psi(\zeta), H_{\rm int} \tilde{S}^{(3)}_{Y_l}\Psi(\zeta))  
 = U \sum_{y \in \partial Y_l}
 \|  c_{y, \uparrow}c_{y, \downarrow} \tilde{S}^{(3)}_{Y_l}\Psi(\zeta) \|^2,
\end{align}
where we define $\partial Y_l=\bar{Y}_l \cap \overline{\bar{Y}_l^c}  \subset \Lambda^\prime$.
Note that for any $y \in \partial Y_l$, there exist
$x_1 \in \Lambda_o \backslash Y_l$ and $x_2 \in Y_l$ such that
$\bar{x_1} \cap \bar{x_2} = {y}$.
Thus, we find a bound
\begin{align}
 &
 \|  c_{y, \uparrow}
 c_{y, \downarrow} \tilde{S}^{(3)}_{Y_l}\Psi(\zeta) \|^2
 \nonumber \\
 &= \sum_{f \in \mathcal{P}}
 \frac{\chi[f(x_1) = y] ~ \chi[f(x_2)=x_2]}
 {(|\xi_{f(x_1),\uparrow}|^2+|\xi_{f(x_1),\downarrow}|^2)
 (|\xi_{f(x_2),\uparrow}|^2+|\xi_{f(x_2),\downarrow}|^2)}
 \prod_{x \in \Lambda_o}
 (|\xi_{f(x),\uparrow}|^2+|\xi_{f(x),\downarrow}|^2)
 \nonumber \\
 &\leq \sum_{f \in \mathcal{P}}
 \frac{1}
 {(|\xi_{f(x_1),\uparrow}|^2+|\xi_{f(x_1),\downarrow}|^2)
 (|\xi_{f(x_2),\uparrow}|^2+|\xi_{f(x_2),\downarrow}|^2)}
 \prod_{x \in \Lambda_o}
 (|\xi_{f(x),\uparrow}|^2+|\xi_{f(x),\downarrow}|^2)
 \nonumber \\
 &\leq \frac{\|\Psi(\zeta)\|^2}{(|q|^{[y]-1-z}+|q|^{-[y]+1+z})^2},
\end{align}
where we have used the indicator function again. For sufficiently large $l$, there exists
$G_2 > 0$ such that
$$
\sum_{y \in \partial Y_l}\frac{1}{(|q|^{[y]-1-z}+|q|^{-[y]+1+z})^2} 
 \leq G_2 l^{d-2}.
$$
Therefore, we have the following inequality
\begin{equation}
 (\delta \tilde{S}_{Y_l}^{(3)} \Psi(\zeta), H \delta \tilde{S}_{Y_l}^{(3)} \Psi(\zeta))
  \leq
  G_2 U l^{d-2} \|\Psi(\zeta)\|^2.
  \label{eq:energy}
\end{equation}
From (\ref{eq:norm}) and (\ref{eq:energy}), we obtain
\begin{equation}
 (\tilde{\Psi}_l(\zeta), H\tilde{\Psi}_l(\zeta)) < \frac{G_2}{G_1} U l^{-1}.
\end{equation}
If we define $F_1 = G_2/G_1$, we find (\ref{eq:gapless2}).

Here, we evaluate norm of $P_0 \tilde{\Psi}_l(\zeta)$. First we consider
inner product between two different ground states
\begin{equation}
 \frac{|(\Psi(\zeta^\prime), \Psi(\zeta))|}
  {\|\Psi(\zeta^\prime)\| \|\Psi(\zeta)\|}.
  \label{eq:innerproduct1}
\end{equation}
From Lemma \ref{lemma:4}, we can represent the inner product in the following
\begin{eqnarray}
\frac{(\Psi(\zeta'), \Psi(\zeta))}{\| \Psi(\zeta') \|\| \Psi(\zeta) \|} =
\frac{\sum_{\varphi \in \mathcal{C}} \prod_{x \in \Lambda_o}
\xi_{\varphi(x)}(\zeta')^*
\xi_{\varphi(x)}(\zeta)}{\sqrt{\sum_{\varphi_1,\varphi_2 \in \mathcal{C}} \prod_{x \in \Lambda_o}
|\xi_{\varphi_1(x)}(\zeta')|^2
|\xi_{\varphi_2(x)}(\zeta)|^2}}.
\label{inner}
\end{eqnarray}
The Schwarz inequality ensures the convergence of the inner product
of the normalized ground state
$$
\frac{|(\Psi(\zeta'), \Psi(\zeta))|}{\| \Psi(\zeta') \|\| \Psi(\zeta) \|} \leq 1.
$$
We estimate this inner product with considering the constraints on the configurations.
We evaluate the product of the norms
\begin{align}
 & \| \Psi(\zeta') \|^2 \| \Psi(\zeta) \|^2
 \nonumber \\
 & = \sum_{f_1,f_2 \in \mathcal{P}} \prod_{x \in \Lambda_o}(|\xi'_{f_1(x),\uparrow}|^2+
 |\xi'_{f_1(x),\downarrow}|^2)
 (|\xi_{f_2(x),\uparrow}|^2+|\xi_{f_2(x),\downarrow}|^2).
 \label{norm}
\end{align}
Here, we abbreviate $\zeta$ and $\zeta'$ by
$$
\xi_{f(x),\sigma}=\xi_{f(x),\sigma}(\zeta), \
\ \ \xi'_{f(x),\sigma}=\xi_{f(x),\sigma}(\zeta').
$$
We evaluate each term in this summation.
A term with arbitrary $f_1,f_2 \in {\mathcal P}$
has a lower bound
\begin{align}
 &\sum_{f_1,f_2 \in {\mathcal P}}\frac{1}{2}
 \left[
 \prod_{x \in \Lambda_o}(|\xi_{f_1(x),\uparrow}'|^2+
 |\xi_{f_1(x),\downarrow}'|^2)
 (|\xi_{f_2(x),\uparrow}|^2+|\xi_{f_2(x),\downarrow}|^2) + 
 (f_1 \leftrightarrow f_2)
 \right] \nonumber \\
 &\geq \sum_{f_1,f_2 \in {\mathcal P}}\sqrt{\prod_{x \in \Lambda_o} (|\xi_{f_1(x),\uparrow}'|^2+
 |\xi_{f_1(x),\downarrow}'|^2)
 (|\xi_{f_2(x),\uparrow}|^2+|\xi_{f_2(x),\downarrow}|^2)\times (f_1 \leftrightarrow f_2)
 } \nonumber \\
 &=\sum_{f_1,f_2 \in {\mathcal P}}\prod_{x \in \Lambda_o}
 R(f_1(x),f_2(x))^2 \times
 \nonumber \\
 & \qquad
 |{\xi'_{f_1(x),\uparrow}}^*\xi_{f_1(x),\uparrow}+
 {\xi'_{f_1(x),\downarrow}}^*\xi_{f_1(x),\downarrow}|
 |\xi'_{f_2(x),\uparrow}{\xi_{f_2(x),\uparrow}}^*+
 \xi'_{f_2(x),\downarrow}{\xi_{f_2(x),\downarrow}}^*|, \nonumber
\end{align}
where a function $R(y_1, y_2)$ is defined for arbitrary $y_1,y_2 \in \Lambda$ by
\begin{equation}
 R(y_1,y_2)^2
  =\frac{\sqrt{
  (|\xi_{y_1,\uparrow}'|^2+|\xi_{y_1,\downarrow}'|^2)
  (|\xi_{y_1,\uparrow}|^2+|\xi_{y_1,\downarrow}|^2)
  \times (y_1 \leftrightarrow y_2)
% (|\xi_{y_2,\uparrow}|^2+|\xi_{y_2,\downarrow}|^2)
% (|\xi_{y_2,\uparrow}'|^2+|\xi_{y_2,\downarrow}'|^2)
  }}{
  |{\xi'_{y_1,\uparrow}}^*\xi_{y_1,\uparrow}+
  {\xi'_{y_1,\downarrow}}^*\xi_{y_1,\downarrow}|
  |\xi'_{y_2,\uparrow}{\xi_{y_2,\uparrow}}^*+
  \xi'_{y_2,\downarrow}{\xi_{y_2,\downarrow}}^*|}.
\end{equation}
In
the practical calculation, we can check
$R(y_1, y_2) > 1$ for arbitrary $y_1,y_2 \in \Lambda$.
Also the Schwarz inequality for the linearly independent
two vectors $(\xi_{y,\uparrow},\xi_{y,\downarrow})$ and $(\xi'_{y,\uparrow},\xi'_{y,\downarrow})$
ensures this relation.
We define a function
$$
R(x) \equiv \min_{f_1,f_2 \in \mathcal{P}} R(f_1(x), f_2(x)),
$$
which is also larger than $1$.
Therefore, we have an upper bound of the inner product between the normalized
ground states
\begin{eqnarray}
\frac{|(\Psi(\zeta'), \Psi(\zeta))|}{\| \Psi(\zeta') \|\| \Psi(\zeta) \|} \leq
\prod_{x \in \Lambda_o} R(x)^{-1}.
\end{eqnarray}
Each factor $R(x)^{-1}$ with a fixed $[x]$ is a constant less than 1,
since the both functions $\xi_{f(x),\sigma}(\zeta)$ and $\xi_{f(x),\sigma}(\zeta')$
depend on $x$ only through $[x]=\sum_{k=1}^d x_j$ for any $f \in \mathcal{P}$.
We define $R = \max_{x \in \Lambda_o} R(x)$, then we find
\begin{equation}
 |(\Psi(\zeta^\prime), \Psi(\zeta))|
  \leq  R^{-L^{d-1}} \| \Psi(\zeta^\prime)\| \|\Psi(\zeta) \|.
  \label{eq:orthogonal1}
\end{equation}
Next, we evaluate an inner product between 
$\delta \tilde{S}_{Y_l}^{(3)} \Psi(\zeta)$ and another ground state.
First, we evaluate 
\begin{align}
 &
 (\Psi(\zeta'), \tilde{S}^{(3)}_{Y_l} \Psi(\zeta))
 \nonumber \\
 & 
 = \frac{1}{2}\sum_{f \in \mathcal{P}} \sum_{x_1 \in Y_l}
 \frac{{\xi'_{f(x_1),\uparrow}}^* \xi_{f(x_1),\uparrow}-
 {\xi'_{f(x_1),\downarrow}}^*
 \xi_{f(x_1),\downarrow}}{{\xi'_{f(x_1),\uparrow}}^* \xi_{f(x_1),\uparrow}+
 {\xi'_{f(x_1),\downarrow}}^*
 \xi_{f(x_1),\downarrow}}
 \prod_{x \in \Lambda_o}({\xi'_{f(x),\uparrow}}^* \xi_{f(x),\uparrow}+
 {\xi'_{f(x),\downarrow}}^*
 \xi_{f(x),\downarrow}).
\end{align}
Then, we have
\begin{align}
 |(\Psi(\zeta'), \tilde{S}^{(3)}_{Y_l} \Psi(\zeta))|
 \leq &
 \frac{1}{2}\sum_{x_1 \in Y_l} 1
 \left|
 \sum_{f \in \mathcal{P}} 
 \prod_{x \in \Lambda_o}({\xi'_{f(x),\uparrow}}^* \xi_{f(x),\uparrow}+
 {\xi'_{f(x),\downarrow}}^*
 \xi_{f(x),\downarrow})
 \right|
 \nonumber \\
 = & \frac{1}{2} G_3 l^d |(\Psi(\zeta'), \Psi(\zeta))|,
\end{align}
where we define $G_3$ by $\sum_{x_1 \in Y_l} 1 = G_3 l^d.$
Also we obtain another upper bound
\begin{equation}
 (\Psi(\zeta), \tilde{S}_{Y_l}^{(3)} \Psi(\zeta))
  \leq
  \frac{1}{2} G_3 l^d \| \Psi(\zeta) \|^2
  .
\end{equation}
Therefore, the inner product between $\delta \tilde{S}_{Y_l}^{(3)} \Psi(\zeta)$ and the
ground state is estimated as
\begin{equation}
 |(\Psi(\zeta'), \delta \tilde{S}^{(3)}_{Y_l} \Psi(\zeta))|
  \leq G_3 l^{d} R^{-L^{d-1}} \|\Psi(\zeta')\| \|\Psi(\zeta))\|.
  \label{eq:orthogonal2}
\end{equation}

Now we estimate $\| P_0 \tilde{\Psi}_l(\zeta)\|$.
Since $\{ \Psi(\zeta_j) \}_{j=0}^{L^d}$ is a complete basis of the ground
states, we can represent $P_0 \tilde{\Psi}_l(\zeta)$ by
\begin{equation}
 P_0 \tilde{\Psi}_l(\zeta)
  = \sum_{j=0}^{L^d} \frac{C_j}{\|\Psi(\zeta_j)\|} \Psi(\zeta_j),
\end{equation}
where $C_j$ is a complex coefficient. Thus, we have
\begin{align}
 \|P_0 \tilde{\Psi}_l(\zeta)\|^2 = &
 \sum_{j=0}^{L^d} \frac{C_j}{\|\Psi(\zeta_j)\|}
 (\tilde{\Psi}_l(\zeta), \Psi(\zeta_j))
 = \sum_{j=0}^{L^d} C_j
 \frac{(\delta \tilde{S}_{Y_l}^{(3)} \Psi(\zeta), \Psi(\zeta_j))}
 {\|\delta \tilde{S}_{Y_l}^{(3)} \Psi(\zeta)\| \|\Psi(\zeta_j)\|}
 \nonumber \\
 < &
 \frac{G_3}{\sqrt{G_1}} l^{(d+1)/2} R^{-L^{d-1}}
 \sum_{j=0}^{L^d} |C_j|, \label{eq:projected1}
\end{align}
where we have used (\ref{eq:orthogonal1}) and (\ref{eq:orthogonal2}).
To evaluate $\sum_j |C_j|$, we consider
\begin{equation}
 \frac{(\Psi(\zeta_j), P_0 \tilde{\Psi}_l(\zeta))}{\| \Psi(\zeta_j) \|}
  =
  \sum_{k=0}^{L^{d}} C_k
  \frac{(\Psi(\zeta_j), \Psi(\zeta_k))}{\|\Psi(\zeta_j)\| \|\Psi(\zeta_k)\|}
  =
  C_j + \sum_{k\neq j} C_k
  \frac{(\Psi(\zeta_j), \Psi(\zeta_k))}{\|\Psi(\zeta_j)\| \|\Psi(\zeta_k)\|}.
\end{equation}
Then, we have
\begin{align}
 |C_j| < &
 \frac{|(\Psi(\zeta_j), P_0 \tilde{\Psi}_l(\zeta))|}{\| \Psi(\zeta_j) \|}
 + \sum_{k\neq j} |C_k|
 \frac{|(\Psi(\zeta_j), \Psi(\zeta_k))|}{\|\Psi(\zeta_j)\| \|\Psi(\zeta_k)\|}
 \nonumber \\
 < &
 \frac{G_3}{\sqrt{G_1}}l^{(d+1)/2} R^{-L^{d-1}}
 + R^{-L^{d-1}} \sum_{k\neq j} |C_k|,
\end{align}
where we have used (\ref{eq:orthogonal1}) and (\ref{eq:orthogonal2}).
If we define $|C_{m}|=\max \{ |C_k| \}_{k=0}^{L^d}$,
then we obtain
\begin{equation}
|C_k| \leq |C_m| <
  \frac{G_3}{\sqrt{G_1}}
  \frac{l^{(d+1)/2} R^{-L^{d-1}}}{1-L^d R^{-L^{d-1}}},
  \label{eq:UB_of_coef}
\end{equation}
for any $k = 0, 1, \cdots, L^d$.
Thus, we obtain
\begin{equation}
 \|P_0 \tilde{\Psi}_l(\zeta)\|^2
  <
  \frac{G_3{^2}}{G_1}
  \frac{(L^d+1)l^{d+1} R^{-2L^{d-1}}}{1-L^d R^{-L^{d-1}}}.
  \label{eq:norm_of_projected}
\end{equation}
from (\ref{eq:projected1}) and (\ref{eq:UB_of_coef}).
If we define $L_1$ by
$1-L_1{}^d R^{-L_1{}^{d-1}} = 1/2$ and $F_2$ by
$F_2 = 2 G_3{}^2/G_1$, then we obtain
\begin{equation}
 \|P_0 \tilde{\Psi}_l(\zeta)\|^2
  <
  F_2
  (L^d+1)l^{d+1} R^{-2L^{d-1}}.
\end{equation}
By definition of the normalized state (\ref{normalized}),
this inequality completes the  
proof of Theorem \ref{th:gapless2}. \qed

\subsection{Proof of Corollary \ref{th:gapless1}}

Here, we prove Corollary \ref{th:gapless1}.
We define a normalized state $\tilde{\Psi}_\perp$ by
\begin{equation}
 \tilde{\Psi}_{\perp} :=
  \frac{(1-P_0) \tilde{\Psi}_L(1)}{\|(1-P_0) \tilde{\Psi}_L(1)\|}.
\end{equation}
An upper bound of $(\tilde{\Psi}_\perp, H \tilde{\Psi}_\perp)$ gives an
upper bound on the lowest excitation energy in a finite system, since
$\tilde{\Psi}_\perp(\zeta)$ is orthogonal to all of the ground state.
Since $H \tilde{\Psi}_L (1) = H (1-P_0) \tilde{\Psi}_L (1)$, the
only remaining task is to estimate $\|(1-P_0) \tilde{\Psi}_L(1)\|$.
From (\ref{eq:norm_of_projected}), we obtain
\begin{align}
 \|(1-P_0) \tilde{\Psi}_L(0)\|^2 = &
 \|\tilde{\Psi}_L(1)\|^2 - \|P_0 \tilde{\Psi}_L(1)\|^2
 <
 1 - \frac{{G_3}^2}{G_1}
 \frac{(L^d+1)L^{d+1} R^{-2L^{d-1}}}{1-L^d R^{-L^{d-1}}}.
\end{align}
If we define $L_2$ by
\begin{equation}
 \frac{{G_3}^2}{G_1}
  \frac{(L_2{}^d+1)L_2{}^{d+1} R^{-2L_2{}^{d-1}}}{1-L_2{}^d R^{-L_2{}^{d-1}}}
  = \frac{1}{2},
\end{equation}
and set $F_3 = 2 F_1$,
then we obtain an upper bound
\begin{equation}
 (\tilde{\Psi}_\perp, H \tilde{\Psi}_\perp)
  < F_3 U L^{-1}.
\end{equation}
This gives an upper bound on the lowest excitation energy in a finite system,
and so completes the proof of Corollary \ref{th:gapless1}.
\qed

\section{Existence of the Spin-Wave Gap \label{sec:SW}}

In this section, we consider our model under the periodic boundary
condition. The properties of ground states and low energy
excitations are very different from a system with the open boundary.
We find only two ground states: the all-spin-up state and the all-spin-down state.
We show that a one-magnon spin-wave excitation 
has an energy gap as in the XXZ model.
The proof is based on Tasaki's argument for the SU(2)
invariant model \cite{Tasaki}. He proved that the one-magnon
spin-wave excitation in the Tasaki model has the same dispersion
relation as that in the ferromagnetic Heisenberg model. These spin-wave
excitations in both models have no energy gap, since they are the
Goldstone mode above the ground states which spontaneously break the
SU(2) spin rotation symmetry.
On the contrary, an energy gap is generated by the anisotropy in our
model as in the XXZ model.
Here, we show only a brief sketch of the proof.

\subsection{Ground States and Spin-Wave Excitation}

First we obtain ground states. We have already found the representation
of a ground state (\ref{eq:rep_of_GS_M}) with the condition
(\ref{eq:cond_for_phi}) in section \ref{sec:GS}. The periodic boundary
condition allows no configuration which satisfies the condition
(\ref{eq:cond_for_phi}) except in the two cases: $\sigma_x = \uparrow$
for all $x \in \Lambda_o$ or $\sigma_x = \downarrow$ for all
$x \in \Lambda_o$. Thus, we conclude that all ground states in the
periodic system are only two fully polarized states $\Phi_\uparrow$ and
$\Phi_{\downarrow}$.

Next we consider the one-magnon spin-wave excitation.
For the spin-wave in our electron model, we consider properties of
the spin-wave state in quantum spin models. The one-magnon spin-wave
state with a wave-number $k \in \K$: $\Phi_{\rm SW}(k)$ satisfies
\begin{equation}
 T_{x} \Phi_{\rm SW}(k) = e^{- i k \cdot x} \Phi_{\rm SW}(k)
  \label{eq:SW_1}
\end{equation}
and
\begin{equation}
 S_{\rm tot}^{(3)} \Phi_{\rm SW}
  = (S_{\rm max} - 1 ) \Phi_{\rm SW},
  \label{eq:SW_2}
\end{equation}
where $x \in \Lambda_o$. The translation operator $T_x$ is
defined by
\begin{equation}
 T_x c_{y, \sigma} T_x^{-1} = c_{x + y, \sigma}
  \quad \mbox{and} \quad
 T_x c_{y, \sigma}^\dagger T_x^{-1} = c_{x + y, \sigma}^\dagger.
\end{equation}
$\K$ is the space of wave-number vectors
\begin{equation}
 \K :=
  \left\{
  \frac{2 \pi n}{L} ~\biggl|~
  n \in {\mathbb Z}^d \cap \left[ -\frac{L-1}{2}, \frac{L-1}{2} \right]^d
  \right\}.
  \label{eq:def_of_K}
\end{equation}
Then, the one-magnon spin-wave state is in the following Hilbert
space ${\cal H}_k$
\begin{equation}
 {\cal H}_k :=
  \left\{
   \Psi \in {\cal H} ~\Bigl|~
   T_x \Psi = e^{-i k \cdot x} \Psi \quad \mbox{and} \quad
   S_{\rm tot}^{(3)} \Psi = \tfrac{1}{2} (|\Lambda_o| - 1) \Psi
  \right\}.
\end{equation}
We define the one-magnon spin-wave state with wave-number $k$ by the lowest
energy state in ${\cal H}_k$. Let $E_{\rm SW}(k)$ be the energy of one-magnon
spin-wave state with wave-number $k$. We can prove the following theorem.

{\theorem(Spin-Wave Gap)
Suppose the $d$-dimensional Hubbard model defined by the Hamiltonian 
(\ref{Hamiltonian}). 
There exist positive constants $t_0, U_0, \lambda_0, C < \infty$ which are 
independent of system volume
such that
\begin{equation}
 \min_{k \in \K} E_{\rm SW}(k) \geq \frac{2 U}{\lambda^{4}}
  \left[
   \frac{d(|q| + |q|^{-1}-2)}{2} - \frac{C}{\lambda}
  \right],
\end{equation}
for $t \geq t_0, U \geq U_0$ and $\lambda \geq \lambda_0$.
\label{th:SW-Gap}}
\vspace*{1em}

This theorem shows that one-magnon spin-wave excitation has a finite gap
for sufficiently large $\lambda$ in a periodic system.

\subsection{Sketch of Proof}

Theorem \ref{th:SW-Gap} is proved by the same approach given in
\cite{Tasaki}. Here, we show only a sketch of the proof.

First we introduce a localized electron operator $a_{x, \sigma}^\dagger$
defined by
\begin{equation}
 a_{x, \sigma}^\dagger := \sum_{ y \in \Lambda}
  \psi_{y, \sigma}^{(x)} c_{y, \sigma}^\dagger,
  \label{eq:def_of_a2}
\end{equation}
where $\psi_{y, \sigma}^{(x)}$ is defined by
\begin{equation}
 \psi_{y, \sigma}^{(x)} :=
   \begin{cases}
    \displaystyle
    \delta_{x, y} - \sum_{j=1}^{d}
    \left(
    \frac{q^{p(\sigma)/4}}{\lambda} \delta_{x - e^{(j)}, y}
    +
    \frac{q^{-p(\sigma)/4}}{\lambda} \delta_{x + e^{(j)}, y}
    \right)
    & \mbox{ if } x \in \Lambda_o \\
    \displaystyle
    \delta_{x, y}    
    + \frac{(q^{-p(\sigma)/4})^\ast}{\lambda} \delta_{x - e^{(j)}, y}
    + \frac{(q^{p(\sigma)/4})^\ast}{\lambda} \delta_{x + e^{(j)}, y}
    & \mbox{ if } x \in \Lambda_j
   \end{cases}
   .
   \label{eq:local_basis}
\end{equation}
The set $\{a_{x,\sigma}^\dag \Phi_{\rm vac}\}_{x \in \Lambda}$
is a basis in the space of single electron states.
We define the dual operator $b_{x, \sigma}$ which
satisfies
\begin{equation}
 \{ b_{x, \sigma}, a_{y, \tau}^\dagger \} = \delta_{x, y}
  \delta_{\sigma, \tau},
  \quad \mbox{and} \quad
 \{ b_{x, \sigma}, b_{y, \tau} \}
 = 0 =
 \{ a_{x, \sigma}^\dagger, a_{y, \tau}^\dagger \}.
 \label{eq:a-com_a-b}
\end{equation}
We represent $b_{x, \sigma}$ in terms of the original electron operator
by
\begin{equation}
 b_{x, \sigma} = \sum_{y \in \Lambda}
  (\tilde{\psi}_{y, \sigma}^{(x)})^\ast c_{y, \sigma}.
  \label{eq:def_of_b}
\end{equation}
Eqs. (\ref{eq:def_of_a2}), (\ref{eq:a-com_a-b}) and (\ref{eq:def_of_b}) mean
\begin{equation}
 \sum_{w \in \Lambda}
  (\tilde{\psi}_{w, \sigma}^{(x)})^\ast \psi_{w, \sigma}^{(y)}
  = \delta_{x, y}
  \quad \mbox{and} \quad
  \sum_{w \in \Lambda}
  (\tilde{\psi}_{x, \sigma}^{(w)})^\ast \psi_{y, \sigma}^{(w)}
  = \delta_{x, y}.
  \label{eq:dual_basis}
\end{equation}
The original electron operators can be written in terms of
$a_{x, \sigma}^\dagger$ and $b_{x, \sigma}$,
\begin{equation}
 c_{x, \sigma}^\dagger
  = \sum_{y \in \Lambda}
  (\tilde{\psi}_{x, \sigma}^{(y)})^\ast a_{y, \sigma}^\dagger
  \quad \mbox{and} \quad
  c_{x, \sigma}
  = \sum_{y \in \Lambda} \psi_{x, \sigma}^{(y)} b_{y, \sigma}.
  \label{eq:c_i.t.o._a_or_b}
\end{equation}
The Hilbert space with $|\Lambda_o|$ electrons is also spanned by the
basis
\begin{equation}
 \left\{
  \left( \prod_{x \in A} a_{x, \uparrow}^\dagger \right)
  \left( \prod_{x \in B} a_{x, \downarrow}^\dagger \right)
  \Phi_{\rm vac}
  \biggl| A, B \subset \Lambda \mbox{ with }
  |A| + |B| = |\Lambda_o|
 \right\},
\end{equation}
because $c_{x, \sigma}^\dagger$ can be written in terms of
$a_{x, \sigma}^\dagger$.

We represent the interaction Hamiltonian in terms of this basis
\begin{equation}
 H_{\rm int} =
  \sum_{x_1, x_2, x_3, x_4 \in \Lambda}
  \left(
   U \sum_{w \in \Lambda}
  (\tilde{\psi}_{w, \uparrow}^{(x_1)}
  \tilde{\psi}_{w, \downarrow}^{(x_2)})^\ast
  \psi_{w, \downarrow}^{(x_3)} \psi_{w, \uparrow}^{(x_4)}
  \right)
  a_{x_1, \uparrow}^\dagger a_{x_2, \downarrow}^\dagger
  b_{x_3, \downarrow} b_{x_4, \uparrow}
  .
  \label{eq:H_int-2}
\end{equation}
It is convenient to introduce a new hopping Hamiltonian
$\tilde{H}_{\rm hop}$ defined by
\begin{equation}
 \tilde{H}_{\rm hop} := t \lambda^2 \sum_{\sigma=\uparrow, \downarrow}
  \sum_{x \in \Lambda^\prime} a_{x, \sigma}^\dagger b_{x, \sigma}
\end{equation}
for the estimation of a lower bound of spin-wave excitation.
$\tilde{H}_{\rm hop}$ satisfies
\begin{equation}
 \tilde{H}_{\rm hop} a_{x, \sigma}^\dagger \Phi_{\rm vac}
  =
  \begin{cases}
   0 & \mbox{if } x \in \Lambda_o \\
   t \lambda^2 & \mbox{if } x \in \Lambda^\prime
  \end{cases}.
\end{equation}
Since $H_{\rm hop} a_{x, \sigma}^\dagger \Phi_{\rm vac}=0$ for
$x \in \Lambda_o$ and $t \lambda^2$ is lowest energy eigenvalue of a
single electron state which is orthogonal to the zero energy states,
then we have $\tilde{H}_{\rm hop} \leq H_{\rm hop}$.

First, we define a basis of ${\cal H}_k$.
To define a convenient basis of ${\cal H}_k$, we define a state
$\Psi_{\mu, A}(k)$ for $\mu=0,1, \cdots, d$ and for a set
$A \subset \Lambda$ with $|A| = |\Lambda_o|-1$ by
\begin{equation}
 \Psi_{\mu, A}(k) :=
  \sum_{w \in \Lambda_o} e^{i k \cdot w} T_w
  a_{e^{(\mu)}, \downarrow}^\dagger
  \left(
   \prod_{v \in A} a_{v, \uparrow}^\dagger
  \right)
  \Phi_{\rm vac},
\end{equation}
where $e^{(\mu)}=o =(0, 0, \cdots, 0)$ for $\mu=0$ and $e^{(\mu)}=e^{(j)}$ for $\mu=j$
$(j=1, 2, \cdots, d)$.
This state satisfies both properties (\ref{eq:SW_1}) and
(\ref{eq:SW_2}). We define another state $\Omega(k)$ by
\begin{equation}
 \Omega(k) = \frac{1}{\alpha(k)}
  \sum_{w \in \Lambda_o} e^{i k \cdot w} T_w
  a_{o, \downarrow}^\dagger b_{o, \uparrow} \Phi_\uparrow
  \propto \Psi_{0, \Lambda_o \backslash \{ o \}},
\end{equation}
which is an approximation of the spin-wave state. We will choose a
constant $\alpha(k)$ in the proof. We define the following basis of
${\cal H}_k$ by
\begin{align}
 \B_{k} :=
 \{ \Omega(k) \} \cup
 \bigl\{ &
 \Psi_{\mu, A}
 ~\bigl|~
 \mu = 0, 1, \cdots, d, \quad A \subset \Lambda
 \nonumber \\
 & 
 \mbox{with } |A| = |\Lambda_o| -1
 \mbox{ and }
 (\mu, A) \neq (0, \Lambda_o \backslash \{ o \})
 \bigr\}.
\end{align}
We define $\tilde{H}$ by $\tilde{H}_{\rm hop} + H_{\rm int}$ and matrix
elements $h[\Phi, \Psi]$ between $\Phi, \Psi \in \B_k$ by the unique expansion
\begin{equation}
 \tilde{H} \Phi = 
  \sum_{\Psi \in \B_{k}} h[\Psi, \Phi] \Psi.
\end{equation}
And we define $D[\Phi]$ by
\begin{equation}
 D[\Phi] := \Re[h[\Phi, \Phi]]
  - \sum_{\Psi \in \B_k \backslash \{ \Phi \}}
  |h[\Phi, \Psi]|.
\end{equation}
Now, we can prove the following lemmas.

{\lemma \label{lemma:SW-base}
Let $E_0(k)$ be the lowest energy eigenvalue of $\tilde{H}$ in the
Hilbert space ${\cal H}_k$. Then, we have
\begin{equation}
 E_0(k) \geq \min_{\Phi \in \B_k} D[\Phi].
\end{equation}
}

{\lemma \label{lemma:SW-LB}
There exist positive constants $t_0, U_0, \lambda_0, C < \infty$ independent of system volume
such that
\begin{equation}
 \min_{\Phi \in \B_{k}} D[\Phi] = D[\Omega(k)]
  \geq
  \frac{2 U}{\lambda^{4}}
  \left[
   \frac{d(|q| + |q|^{-1})}{2}
   - \sum_{j = 1}^{d} \cos \left( 2 k \cdot e^{(j)} + \theta \right)
   - \frac{C}{\lambda}
  \right]
\end{equation}
for $t \geq t_0$, $\lambda \geq \lambda_0$ and $U \geq U_0$.
}

\vspace*{1em}

We find the proof of Lemma \ref{lemma:SW-base} in subsection 6.1 of
ref. \cite{Tasaki}. Lemma \ref{lemma:SW-LB} is obtained by a direct
evaluation.

Now, we can prove Theorem \ref{th:SW-Gap}.
Since $\tilde{H} \leq H$, then we have $E_0(k) \leq E_{\rm SW}(k)$.
Thus we find
\begin{align}
 \Delta E \geq \min_{k \in {\cal K}} E_0(k)
 \geq \min_{k \in {\cal K}}
  \frac{2 U}{\lambda^{4}}
  \left[
   \frac{d(|q| + |q|^{-1})}{2}
   - \sum_{j = 1}^{d} \cos \left( 2 k \cdot e^{(j)} + \theta \right)
   - \frac{C}{\lambda}
  \right]
\end{align}
from Lemma \ref{lemma:SW-base} and \ref{lemma:SW-LB}.
This concludes Theorem \ref{th:SW-Gap}. \qed

\section{Summary\label{sec:summary}}

In this paper, we construct a set of exact ground state with a
ferromagnetic domain wall structure and a spiral structure in a deformed
flat-band Hubbard model under an open boundary condition.
We have studied excited states
above the domain wall ground state. There exists a gapless
excitation 
above the domain wall ground state in dimensions higher than
one. This excited state is constructed by acting a local operator 
near the domain wall on the ground state.
We study this model also under the periodic boundary condition.
In this case a ground state becomes the all-spin-up or -down state. 
We have shown the energy gap of the spin-wave excitation above
the all-spin-up ground state. 
 These properties of the excitations above the ground states
are similar to those in the XXZ model.\\

\noindent{\large\bf Acknowlegements}\vspace*{1em}

We would like to thank T. Koma, B. Nachtergaele, A. Tanaka and H. Tasaki
for helpful suggestions and great encouragements. We are grateful to
T. Asaga and T. Fujita for kindly reading the manuscript.

\end{document}